\documentclass[prb,superscriptaddress,twocolumn]{revtex4-1}
\usepackage[american,british]{babel}
\usepackage[pdftex]{graphicx} 
\usepackage{graphicx, xcolor}
\usepackage{amsmath,amsthm,amssymb}
\usepackage{amsmath}
\usepackage{color}

\definecolor{darkGreen}{RGB}{0,110,0}
\definecolor{darkBlue}{RGB}{0,0,130}
\usepackage[colorlinks,citecolor=darkGreen,linkcolor=darkBlue,urlcolor=blue,hyperindex]{hyperref}

\newcommand{\bra}[1]{\left\langle #1 \right|}
\newcommand{\ket}[1]{\left| #1 \right\rangle}

\newcommand{\mytitle}{Signatures of many-body localisation in the dynamics of two-site entanglement}

\begin{document}

\title{\mytitle}

\author{Fernando Iemini}
\affiliation{Abdus Salam ICTP, Strada Costiera 11, I-34151 Trieste, Italy}

\author{Angelo Russomanno}
\affiliation{Abdus Salam ICTP, Strada Costiera 11, I-34151 Trieste, Italy}
\affiliation{NEST, Scuola Normale Superiore \& Istituto Nanoscienze-CNR, I-56126 Pisa, Italy}

\author{Davide Rossini}
\affiliation{NEST, Scuola Normale Superiore \& Istituto Nanoscienze-CNR, I-56126 Pisa, Italy}

\author{Antonello Scardicchio}
\affiliation{Abdus Salam ICTP, Strada Costiera 11, I-34151 Trieste, Italy}
\affiliation{INFN, Sezione di Trieste, Via Valerio 2, I-34127 Trieste, Italy}

\author{Rosario Fazio}
\affiliation{Abdus Salam ICTP, Strada Costiera 11, I-34151 Trieste, Italy}
\affiliation{NEST, Scuola Normale Superiore \& Istituto Nanoscienze-CNR, I-56126 Pisa, Italy}

\date{\today}

\begin{abstract}
  We are able to detect clear signatures of dephasing -- a distinct trait of many-body 
  localisation (MBL) -- via the dynamics of two-site entanglement, quantified through the concurrence. 
  Using the protocol implemented by M.~Schreiber {\it et al.} [Science {\bf 349}, 842 (2015)], 
  we show that in the MBL phase the average two-site entanglement decays in time as a power law, 
  while in the Anderson localised phase it tends to a plateau. 
  The power-law exponent is not universal and displays a clear dependence on the interaction strength. 
  This behaviour is also qualitatively different from the ergodic phase, 
  where the two-site entanglement decays exponentially. All the results have been obtained by means 
  of time-dependent density matrix renormalisation group simulations, and further corroborated 
  by analytical calculations on an effective model. Two-site entanglement has been already measured 
  in cold atoms: our analysis paves the way for the first direct experimental test 
  of many-body dephasing in the MBL phase. 
\end{abstract}


\maketitle

\section{Introduction}
\label{intro}
The phenomenon of many-body localisation~\cite{basko2006metal,oganesyan2007localization,Nandkishore:2015aa} 
(MBL) refers to the breakdown of ergodicity in generic disordered many-body systems, due to quantum effects. 
This is a striking counterexample to the fundamental assumptions of statistical mechanics about 
the thermalisation of an isolated system. For any non-integrable classical many-body Hamiltonian system, 
the dynamics is ergodic in phase space, eventually leading to thermalisation. This occurs 
even for systems close to integrability via Arnold diffusion, a phenomenon strictly related 
to the celebrated KAM theorem~\cite{Berry_LH84,Arnold:book}.
In quantum systems there is a striking exception: destructive interference between matter waves 
forbids a system in the MBL phase to thermalise. 
Quantum effects make the system non-ergodic: no part of it acts as reservoir for the rest of the system. 

At first glance, the existence of MBL is astonishing: due to the presence of interactions, one expects 
the quantum system to be non-integrable and to display ergodicity and thermalisation~\cite{Polkovnikov_RMP11}. 
This behaviour is however strange only apparently: indeed a system in the MBL phase 
can be mapped into an integrable system with an extensive number of local integrals 
of motion~\cite{Abanin.2103.LIOM,ros2015integrals,Abanin_PRB,imbrie2014many}. 
Traditionally, integrable systems are isolated points in the space of Hamiltonians, 
both from a classical~\cite{Berry_LH84} and a quantum~\cite{Pasqualazzo} perspective; 
on the opposite, in MBL, integrability and non-ergodicity do not require any fine tuning. 
Remarkably, MBL has been recently conjectured to occur even in systems 
without disorder~\cite{Mauro_PRB,Pino_arXiv15}: the contrast with the behaviour 
of classical systems is even more striking.

In some sense, the MBL phase is the continuation of the Anderson localised (AL) 
phase~\cite{anderson1958absence,Abrahams2010} of non-interacting particles, when interactions are turned on: 
the two phases share several properties, mainly the absence of transport of any physical quantity. 
At the same time MBL has distinct features that make it qualitatively different from Anderson localisation. 
From one hand, while transport is frozen, correlations can still propagate in the MBL phase. 
This gives rise to a non-trivial dynamics of entanglement which is absent in the AL phase 
and which we will better discuss later.
From the other hand, the transition to the MBL phase does not emerge in thermodynamic quantities, 
but rather in transport and time-correlation functions. This is indeed a \emph{dynamical} transition, 
which requires appropriate observables in order to be identified. 
These very special properties have been recognised thanks to a constantly growing theoretical activity, 
whose aims are elucidating the distinguishing features of MBL and finding ways to detect them 
in the experiments. 

Several works have characterised the MBL phase by: the absence of transport of charge, spin, 
mass~\cite{basko2006metal,pal2010mb,Luitz:2016aa,PhysRevLett.117.040601} 
or energy~\cite{Kerala-Varma:2015aa}; an emergent robust 
integrability~\cite{huse2014phenomenology,Abanin.2103.LIOM,imbrie2014many,ros2015integrals,Abanin_PRB}; 
a logarithmically slow but unbounded growth of entanglement~\cite{znidaric2008many,Bardason2012,Abanin.2013.EntGrowth}; 
a peculiarly sparse structure of eigenfunctions~\cite{de2013ergodicity,Monthus:2016aa}; 
the behaviour of observables after a quantum quench~\cite{Canovi_PRB11,Abanin.2014}; 
the persistence of area law for entanglement up to arbitrary temperatures and for eigenstates 
of arbitrary energy in the spectrum~\cite{Nandkishore:2015aa,berkone,belone,pal2010mb}; 
the ability to protect discrete symmetries~\cite{huse2013localization} even at infinite temperature. 
A comprehensive description of this activity 
can be found in the reviews~[\onlinecite{Nandkishore:2015aa,altman-review}]. 
At the same time several different proposals have been put forward in order to experimentally detect MBL. 
We quote, for example, the interferometric probe based on coherent spin 
manipulations~\cite{Demler-interferometry}, the search for revivals 
of the magnetisation~\cite{Moore-revivals} or the temporal fluctuations around stationary values 
of local observables~\cite{Abanin.2014}.

The intense theoretical efforts of the last decade stimulated an exciting race towards 
its experimental verification. Last year the first beautiful experiments providing evidence 
of MBL appeared in cold atomic systems~\cite{schreiber2015observation,schn_MBL_per} 
and trapped ions~\cite{Monroe_MBL}.
However, it is still debatable whether unique features of MBL, which are not present in AL systems, 
have been observed or not: experiments have focused on the propagation of particles 
which are frozen in both phases. 
It would be highly desirable to have a direct experimental test discriminating between these two cases, 
in order to probe the MBL dephasing mechanism. 
From a theoretical perspective, several different observables/protocols have been proposed 
to this aim (see above), but, in many cases, they are difficult to be implemented experimentally. 
The purpose of this paper is to overcome these difficulties: we analyse in detail a probe of MBL, 
which is able to discriminate it from the AL phase and is experimentally accessible 
within the existing technology. We are going to show that this probe is the two-site entanglement.

The dynamics of two-site entanglement has been recently measured in optical lattices and in trapped ions, 
respectively in Refs.~[\onlinecite{Fukuhara_ent}] and~[\onlinecite{Jurcevic_ent}].
The experiment of Fukuhara {\it et al.}~\cite{Fukuhara_ent} considers a system of atoms governed 
by a Bose-Hubbard Hamiltonian. The spins of atoms are initially in a ferromagnetic phase: 
after flipping a spin at a given site (local quench protocol), the entanglement between 
neighbouring spins is measured as a function of time. 
As we are going to discuss in detail, we consider a slight modification of this quench protocol, 
namely the one implemented in Ref.~[\onlinecite{schreiber2015observation}].
Using this approach, and measuring both the imbalance and the two-site entanglement, we are able 
to extract the key properties of MBL phase. In particular, we can highlight clear differences 
with the AL phase in the measure of the two-site entanglement.

Entanglement plays an important role in MBL. 
While transport (of energy, spin, mass or other macroscopically conserved quantities) is frozen 
both in MBL and in AL phases, quantum correlations can still propagate in the MBL phase, 
giving rise to entanglement between distant sites of the system. 
In this context, the mapping of any MBL system to an integrable system with an extensive number 
of local integrals of motion is crucial. Thanks to this mapping, even in the absence of transport, 
when {\em populations} in every site are stationary, it can be shown that {\em coherences} 
of distant sites evolve in a non-trivial way (more details are in Sec.~\ref{lbit-model}). 
This phenomenon is defined as many-body dephasing: it is ultimately responsible for the unbounded 
(but slow) growth of the entanglement entropy. The situation in the AL phase is very different. 
In this case, the propagation of correlations and the entanglement growth stop after a while. 

Several studies (see, e.g., Refs.~[\onlinecite{Dechiara_JSTAT06,znidaric2008many,Bardason2012,Abanin.2013.EntGrowth,Deng_arX16, Roosz2014, Igloi2012, Zhao2016}]) 
have analysed the evolution of the entanglement entropy of large blocks in disordered spin chains. 
Its logarithmic growth~\cite{znidaric2008many,Bardason2012,Abanin.2013.EntGrowth}, 
intimately related to the existence of an extensive number of local integrals of motions, 
has been identified as a unique trait of MBL. However, despite recent very interesting 
progresses~\cite{Greiner_ent}, the entanglement block-entropy is very hard to be measured in a many-body 
context (virtually impossible on increasing the block size). On the opposite, 
the two-site entanglement we are considering here is directly accessible in cold-atom experiments. 

This paper is organised as follows. In the next section we introduce the model and the quench protocol 
we are going to simulate. Both are chosen to be essentially identical to those implemented 
in the experiment of Ref.~[\onlinecite{schreiber2015observation}]. 
The two-site entanglement will be quantified through the concurrence defined in Sec.~\ref{concu}. 
Section~\ref{results} contains the results of our density matrix renormalisation group simulations. 
They show that the MBL phase is characterised by a typical power-law decay of the concurrence. 
This behaviour strongly contrasts with AL, where the concurrence reaches a non-vanishing stationary value; 
it is also very different from the ergodic phase, where the concurrence abruptly vanishes after a 
short transient. We are able to study how the AL phase is reached as a vanishing-interaction limit 
of the MBL: the power-law decay of the concurrence starts after a stationary, metastable, plateau 
whose extension in time diverges as a power law when the interaction vanishes. 
In Sec.~\ref{lbit-model}, we show that this power-law behaviour is reproduced 
by a phenomenological integrable model of interacting qubits (the so-called ``$\ell$-bit model''): 
this agrees with the fact that our MBL system can be mapped into an integrable system. 
In Sec.~\ref{bound}, we discuss a number of additional effects (the role of number fluctuations, 
finite temperature, control of laser pulses) that may arise when measuring entanglement 
from experimental data. We also provide a (more easily measurable) bound to the concurrence,
which gives very accurate results and faithfully reproduces the essential phenomenology. 
Finally, Sec.~\ref{conclusions} is devoted to our conclusions and perspectives for future work.

\section{The Model}
\label{model}
We consider a generalisation of the Aubry-Andr\'e model~\cite{Aubry-Andre}, which can be realised 
be means of a two-species Bose-Hubbard model in the presence of a periodic potential 
incommensurate with the lattice spacing. This kind of potential is also defined as quasi-periodic, 
pseudo-random or Aubry-Andr\'e potential. In the limit in which the on-site interaction is dominant 
with respect to the hopping, fluctuations in the number of particles on each site are frozen. 
It is then possible to derive an effective Hamiltonian in the subspace where the occupation is fixed
to one particle per site. Here the dynamics is governed by 
a spin-1/2 XXZ Hamiltonian~\cite{kuklov_XXZ,duan_XXZ} where the two eigenstates of 
$\hat{S}_j^z$ represent the occupation of the $j$th site by one of the two species
(here $\hat{S}_j^\alpha \equiv \hat \sigma^\alpha_j/2$, where $\hat \sigma^\alpha_j$ are
the usual Pauli matrices on site $j$, with $\alpha=x,y,z$).
In the presence of an Aubry-Andr\'e potential, the effective spin model also includes 
an inhomogeneous magnetic field leading to the Hamiltonian
\begin{eqnarray} 
  \hat{H} = & - & \sum_j \left[ J (\hat{S}_j^+ \hat{S}_{j+1}^- + {\rm H.c.}) 
    + V \hat{S}_j^z \hat{S}_{i+1}^z \right] \nonumber \\
  & + & \Delta \sum_j \big[ \cos(2\pi \beta j + \phi) \hat{S}_j^z \big] \, ,
\label{hamiltonian}
\end{eqnarray}
where $\hat{S}_j^{\pm} = \hat S^x_j \pm i \hat S^y_j$ are the raising/lowering operators. 
The third term in Eq.~\eqref{hamiltonian} is due to the external quasi-periodic on-site potential: 
the coupling strength appears as the parameter $\Delta$, the inverse of the incommensurate wavelength 
as an irrational number $\beta$, and $\phi$ is a phase. 
In the following, we will consider a one-dimensional optical lattice with $L$ sites
and open boundary conditions. 

Equation~\eqref{hamiltonian} can be mapped, via a Jordan-Wigner transformation, 
onto a spinless fermionic Hubbard model with a quasi-periodic Aubry-Andr\'e chemical potential:
\begin{eqnarray}
  \hat{H} = & - & \sum_j \left[ J (\hat a^\dagger_j \hat a_{j+1} + {\rm H.c.}) 
    + V \hat{n}_j \hat{n}_{j+1} \right] \nonumber \\
  & + & \Delta \sum_j \big[ \cos(2\pi \beta j + \phi) \hat{n}_j \big] \, ,
  \label{fermion.model}
\end{eqnarray}
where $\hat a^{(\dagger)}_j$ is the annihilation (creation) fermion operator, 
$\hat{n}_j = \hat a^\dagger_j a_j$ the local number operator, $J$ the tunneling matrix element 
between neighbouring sites, and $V$ the nearest-neighbour interaction. 
The existence of a MBL phase in this model was rigorously established in Ref.~[\onlinecite{Mastropietro_PRL}].

In the non-interacting case ($V=0$), when the amplitude $\Delta$ of the quasi-periodic potential 
overcomes the threshold $\Delta_c=2$, the system undergoes a transition from an ergodic 
to an AL phase~\cite{Aubry-Andre}. This transition has been recently observed in a cold-atom
experiment~\cite{roati2008anderson}.
The interacting case ($V \ne 0$) presents three distinct phases, depending on the choice 
of the coupling constants: ergodic, MBL, and AL phases. Although an accurate analysis 
of the phase diagram is not the aim of this work, we identified the parameters leading to 
the different phases, without however dwelling on the precise location of the phase boundaries. 
The resulting (approximate) phase diagram in the $\Delta$-$V$ plane is reported in Appendix~\ref{diagram}.

In order to detect signatures of MBL in the two-site entanglement, we study its time-dependence 
after a quantum quench. The protocol we consider is the same as 
in Ref.~[\onlinecite{schreiber2015observation}]. We initialise the system in the N\'eel state
\begin{equation}
  | \psi (t=0) \rangle = \mid \uparrow,\downarrow, \cdots, \uparrow,\downarrow\rangle\,.
\label{initial}
\end{equation}
Then we follow the time evolution governed by Hamiltonian~\eqref{hamiltonian}, 
working in the subspace with total conserved spin 
$S^z_{\rm tot} \equiv \sum_j \langle \hat S^z_j \rangle = 0$. 
We average the quantities of interest over many realisations of pseudo-disorder, 
through a random sampling of the phase $\phi$ in the on-site potential. 
Through the rest of the paper, we will fix the inverse wavelength of the quasi-random potential 
to $\beta = 532/738$. We make this choice because this is the better approximation 
to an irrational number which can be done in the experiments (it is the one used 
in Ref.~[\onlinecite{schreiber2015observation}]). We further set $J=1$ and $\hbar =1$.

\section{Two-site entanglement and concurrence}
\label{concu}
In the model of Eq.~\eqref{hamiltonian}, the entanglement between two-sites can be quantified 
through the concurrence~\cite{Woot_PRL}. Let us consider two sites, $i$ and $j$, and define 
$\rho_{i,j}$ as the reduced density matrix describing the subsystem formed by these two sites. 
The concurrence $C_{i,j}$ measures the entanglement between the two spins located at $i$ and $j$,
minimised over all the possible decompositions of the matrix
\begin{equation}
  \rho_{i,j} = \sum_{a=1}^4 p_a \ket{\psi_a}\bra{\psi_a} \,,
\end{equation}
with arbitrary states $|\psi_a\rangle$ and $\sum_a p_a=1$ (with $p_a\geq 0$).
With this definition, it can be shown that~\cite{Woot_PRL}: 
$C_{i,j}=\max\{0,\lambda^{(1)} - \lambda^{(2)} - \lambda^{(3)} - \lambda^{(4)}\}$, 
where $\lambda^{(\alpha)}$ are the square roots of the eigenvalues of the product matrix 
$R=\rho_{i,j} \tilde{\rho}_{i,j}$, taken in descending order. The spin flipped matrix $\tilde{\rho}$ 
is defined as $\tilde{\rho} \equiv (\sigma^y \otimes \sigma^y) \rho^* (\sigma^y \otimes \sigma^y)$,
where the complex conjugate is taken in the standard basis.
If $C_{i,j}=0$, then there is a decomposition of the reduced density matrix $\rho_{i,j}$ 
in which all the states $\ket{\psi_a}$ are separable.
The concurrence has been employed several times to analyse many-body systems 
(see Ref.~[\onlinecite{amico2008entanglement}] for a review). Here we are going to show that 
its dynamics is able to distinguish between ergodic, MBL, and AL phases. 

In the case we are considering, the total magnetisation $S^z_{\rm tot}$ along the $z$-axis is conserved: 
if we express $\rho_{i,j}$ in the $z$-basis, we find a particularly simple block-diagonal form
\begin{equation}
  \rho_{i,j} = \left( \begin{array}{cccc}
    P_{\uparrow \uparrow} & 0 & 0 & 0 \\
    0 & P_{\uparrow \downarrow} & \rho_{\uparrow \downarrow} & 0 \\
    0 & \rho_{\uparrow \downarrow}^* & P_{\downarrow \uparrow} & 0 \\
    0 & 0 & 0 & P_{\downarrow \downarrow} 
  \end{array} \right) \,,
\end{equation}
where 
\begin{eqnarray}
  P_{\uparrow \uparrow} &=& \langle (\tfrac12 + \hat S^z_i) (\tfrac12 + \hat S^z_j)\rangle \,, \nonumber \\
  P_{\uparrow \downarrow} &=& \langle (\tfrac12 + \hat S^z_i) (\tfrac12 - \hat S^z_j) \rangle \,, \nonumber \\
  P_{\downarrow \uparrow} &=& \langle (\tfrac12 - \hat S^z_i) (\tfrac12 + \hat S^z_j) \rangle \,, \nonumber \\
  P_{\downarrow \downarrow} &=& \langle (\tfrac12 - \hat S^z_i) (\tfrac12 - \hat S^z_j) \rangle \,, \nonumber \\
  \rho_{\uparrow \downarrow} &=& \langle \hat S_i^x \hat S_j^x + \hat S_i^y \hat S_j^y +i [\hat S_i^y \hat S_j^x - \hat S_j^x \hat S_j^y] \rangle \nonumber
  \label{spin-spin}
\end{eqnarray}
(analogous expressions can be written in the fermionic representation).
The concurrence can be analytically computed in this case: 
\begin{equation} \label{conc0:eqn}
  C_{i,j} = 2 \max\left[ 0,|\rho_{\uparrow \downarrow}| - \sqrt{P_{\uparrow \uparrow} P_{\downarrow \downarrow} }\right]\,,
\end{equation}
thus reducing to a very simple form which only contains $zz$ expectation values and a
term $|\rho_{\uparrow \downarrow}|$ which is proportional to the local spin current 
(and vanishes in the long-time limit, due to localisation). 
It is important to keep in mind this observation, in view of the 
analysis that will be presented in Sec.~\ref{bound}.

Because of the incommensurate potential, the concurrence (as well as other observables) 
will be site-dependent. To overcome this difficulty, we choose to analyse a single expression 
containing information on all the pairs of sites; namely, the square of the quasi-disorder 
average concurrence summed over the sites:
\begin{equation} 
  \mathcal{C}(t) = \sum_{i,j \in \mbox{bulk}} \Big[ \overline{C_{i,j}(t)} \Big]^2 \, .
  \label{conc:eqn}
\end{equation}
The bar indicates the quasi-disorder average which is performed over different 
realisations of $\phi$; in order to avoid finite-size effects due to the edges, 
we restrict the summation over $i$ and $j$ to the bulk. Precisely, we consider $L/3\leq i,j \leq 2L/3$. 
The quantity defined in Eq.~\eqref{conc:eqn} allows us to discuss in a compact way the results 
for the two-site entanglement. Moreover (together with the 1-tangle) it allows to extract 
information on the residual multi-partite entanglement of the two selected 
sites~\cite{Coffman_tangle,Osborne_tangle}. The behaviour of $\mathcal{C}(t)$
also reflects the so-called monogamy properties of entanglement:
A given spin cannot be highly entangled with more than one other spin in the system. 
We will see that monogamy is useful to understand the results of this paper.
For simplicity, in the rest of the presentation we will refer to the quantity $\mathcal{C}(t)$ 
of Eq.~\eqref{conc:eqn} as the concurrence.

The two-site entanglement is contained in the reduced density matrix $\rho_{i,j}$, and consequently 
can be expressed through the different spin-spin correlations [see Eqs.~\eqref{spin-spin}]. 
Temporal fluctuations around the stationary values of local, as well as two-spin observables, 
have been shown to decay as power laws~\cite{Abanin.2014}.
Despite providing a good insight about the dynamics, a direct relation between the behaviour 
of the fluctuations and the concurrence cannot be drawn because entanglement results 
in a complicate function of the correlators. In general it has been shown that, in most cases, 
the two-site entanglement is not directly related to the properties 
of correlation functions~\cite{amico2008entanglement}. 

Complementary to the entanglement analysis, we also study the time evolution of the imbalance 
in the occupation between even ($e$) and odd ($o$) sites~\cite{schreiber2015observation}. 
In the particle representation, this is the difference in the occupation of even and odd sites
$\mathcal{I} = (N_e - N_o)/(N_e+N_o)$; in the spin representation it is defined as 
\begin{equation}
  \mathcal{I} = \frac{\langle S^z_e \rangle - \langle S^z_o \rangle}{1+ \langle S^z_e \rangle + \langle S^z_o \rangle}\,.
  \label{imbalance}
\end{equation}
The imbalance has been measured experimentally~\cite{schreiber2015observation,schn_MBL_per}: 
it has been observed that it tends asymptotically to a non-vanishing stationary value, 
both in the MBL and AL phases. As we are going to show, by analysing the time-dependence 
of $\mathcal{I}$ and of $\mathcal{C}$, we can distinguish between the three different phases. 
More importantly, we are also able to capture the subtle dephasing mechanisms occurring in the MBL phase.

\section{Results}
\label{results}
This section will be entirely devoted to the discussion of the outcomes of our simulations, 
concerning the dynamics of the concurrence and the imbalance. In all the cases discussed here, 
we initially prepare the system in the state $\ket{\psi(t=0)}$ [see Eq.~\eqref{initial}] 
and we study its subsequent evolution.
We choose Eq.~\eqref{initial} as it is most relevant for an experimental
 verification of our results.
The choice of a different initial state should not present qualitative
changes in the dynamics of our observables in the localised phase, as will become clear
from the analysis of a phenomenological model in Sec.\ref{lbit-model}. 
In order to have a better comparison with the existing results on the relation 
between many-body dephasing and growth of entanglement, we will also compare our results 
with the behaviour of the entropy. 
The message we would like to convey is that signatures of MBL that emerge in 
the entanglement entropy are evident also in the two-site entanglement, but the last object 
has the important advantage of being easier to access in experiments.

\begin{figure}
  \includegraphics[scale=0.45]{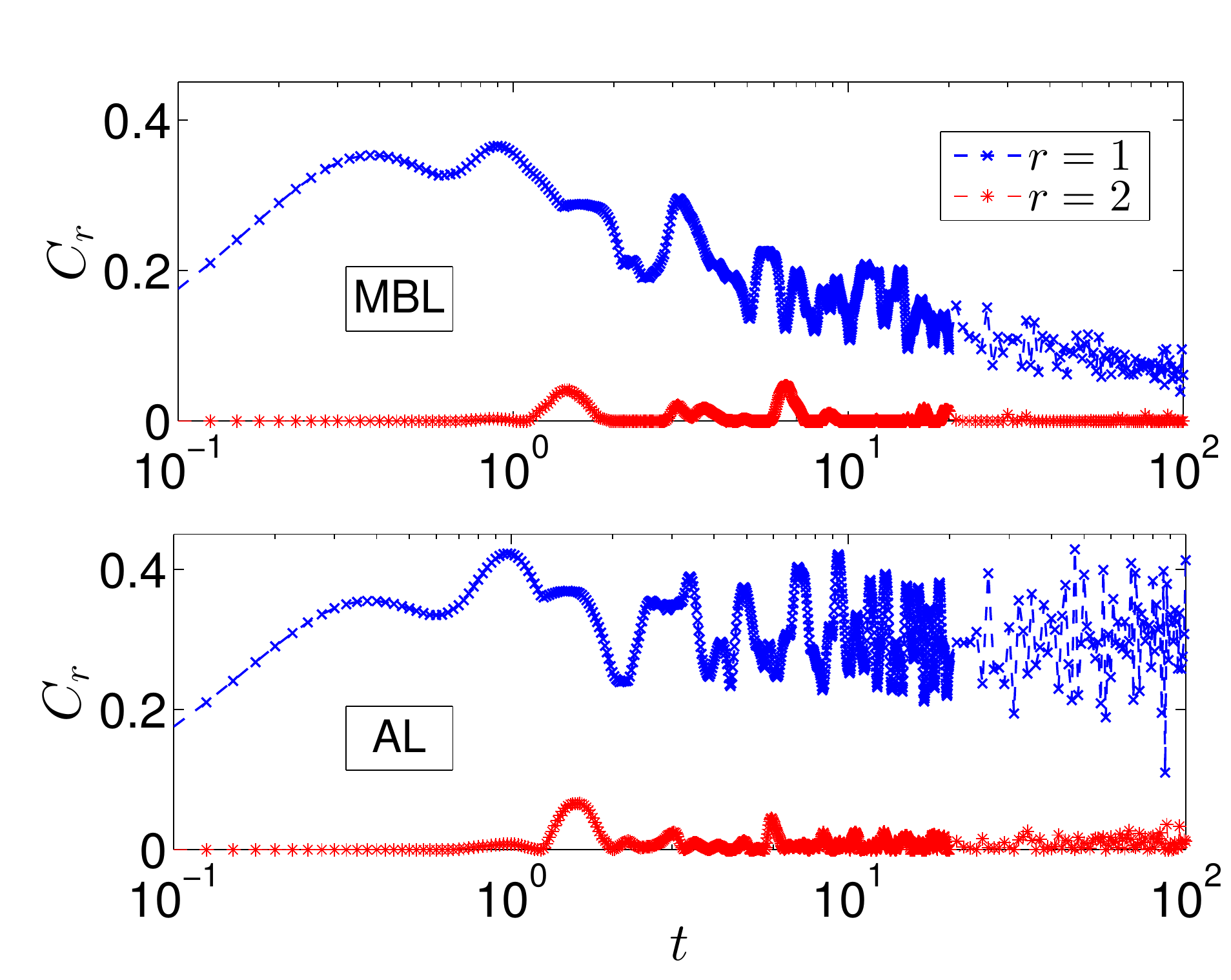}
  \caption{Time evolution for the concurrence at distinct $r = |i-j|$ distances, 
    in a system with $L=24$ sites, averaged over $16$ pseudo-disorder realisations. 
    We consider sites at the center of the lattice, precisely, $r=1 \leftrightarrow (i=L/2, j=L/2+1)$ 
    and $r=2 \leftrightarrow (i=L/2-1, j=L/2+1)$. The top panel refers to the MBL phase, 
    the bottom panel to	the AL phase. In the ergodic phase the concurrence for $r=2$ 
    is several order of magnitude smaller than $C_{r=1}$, and could not be distinguished from zero 
    in the scale of the plot. The results are qualitatively similar if we consider different sites 
    in the bulk. The parameters are $\Delta = 3$, $V = 1$ for the top panel 
    and $\Delta = 4$, $V = 0$ for the bottom panel.}
  \label{fig:entXdist}
\end{figure} 

The dynamics of model~\eqref{hamiltonian} has been simulated using the time-evolving 
block decimation (TEBD) strategy on matrix product states~\cite{Daley_JSTAT04,Schollwock_rev}. 
We used a time step $\Delta t \leq 0.1$ (depending on the model parameters), 
a maximum bond dimension $m=200$, and a Trotter order equal to $4$, leading to negligible error 
thresholds for all the observables under analysis. In the specific case of non-interacting systems, 
however, we have evolved the state in time using the covariance matrix 
(to make use of the simplifications arising for quadratic Hamiltonians).
We have also considered systems with different size, ranging from $L = 12$ to $L=30$
  for the interacting model ($V \neq 0$), and up to $L=240$ for the non-interacting model ($V=0$).
  We carefully verified that all the data presented below are robust with $L$
  (provided $L \gtrsim 20$), and thus our claims do not suffer 
  appreciable finite-size corrections.
  Further details on this issue are provided in Appendix~\ref{finitesize}.

Since the definition of Eq.~\eqref{conc:eqn} involves a summation over many lattice points, 
it is useful to understand whether there are dominant contributions to the sum. 
This analysis is reported in Fig.~\ref{fig:entXdist}, where the concurrence is plotted as a function 
of time for nearest-neighbour and next-nearest-neighbour lattice sites. 
The coupling constants are chosen in such a way to be in the AL (bottom) or MBL (top) phase. 
We did not plot similar curves for the ergodic phase as the next-nearest-neighbour concurrence 
could not be distinguished from zero on the scale of the plot. In all the phases, the concurrence 
between sites distant more than two lattice constants is essentially negligible or vanishing. 
We believe that this behaviour could be indirectly linked to the properties of the eigenstates 
of Hamiltonian~\eqref{hamiltonian}, which display an exponential decay of the concurrence 
with the distance between sites~\cite{Bera_PRB}. 
Therefore, although we use the definition in Eq.~\eqref{conc:eqn}, it is useful to keep in mind 
that the results we are going to present in the rest of the section
essentially reflect the behaviour of the nearest-neighbour concurrence. 
We also remark that our results are robust with
  respect to pseudo-disorder averages, in the sense that the error induced by such averaging
  is barely visible on the scale of the various figures,
  and does not affect our conclusions (see Appendix~\ref{finitesize}
  for a more detailed discussion).
  
In the following subsections we will discuss in detail the dynamics of our system 
in the AL, MBL and ergodic phases.

\subsection{Anderson localised phase}
\begin{figure}
  \includegraphics[width=8.2cm]{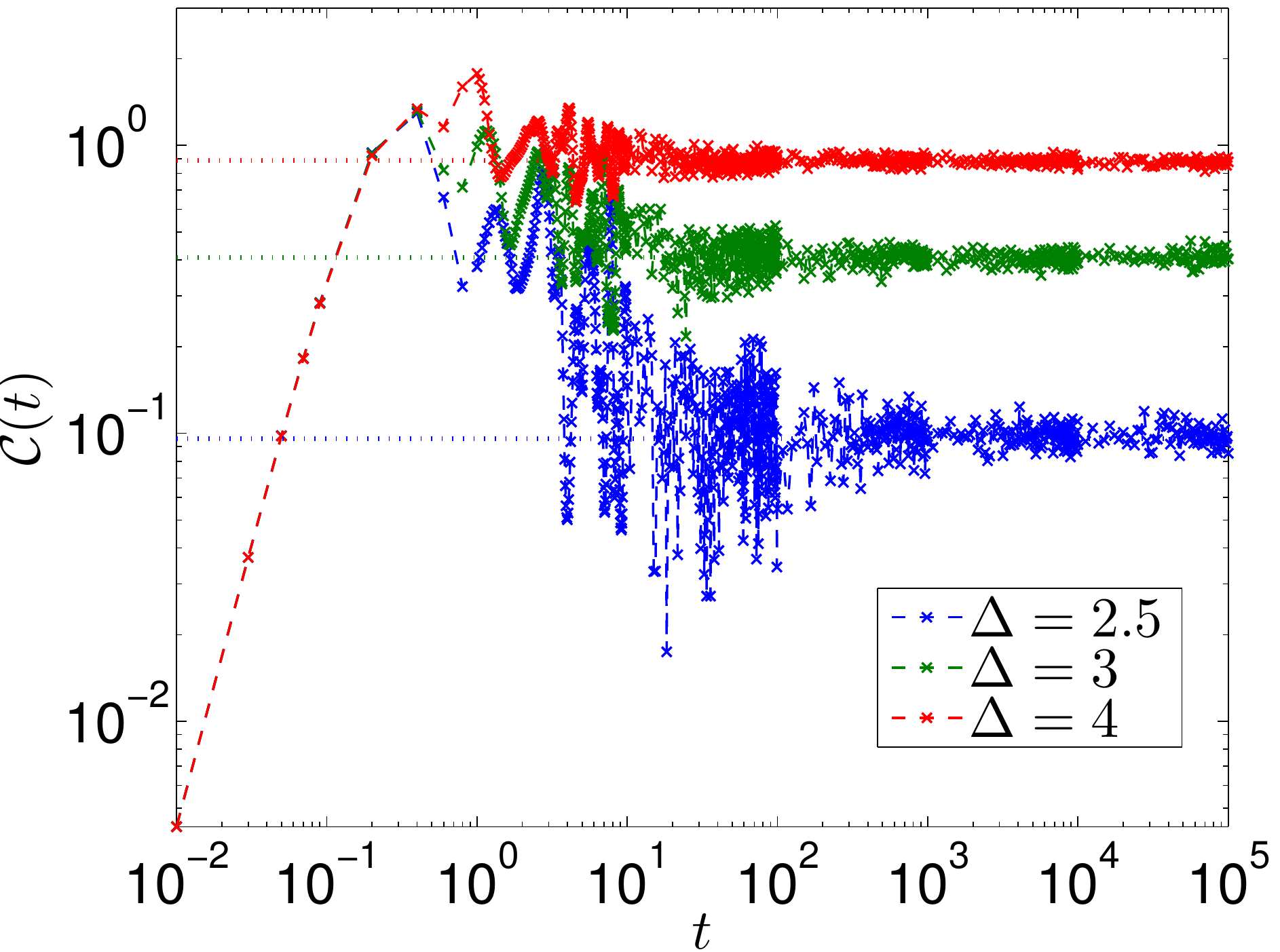}
  \includegraphics[width=8.2cm]{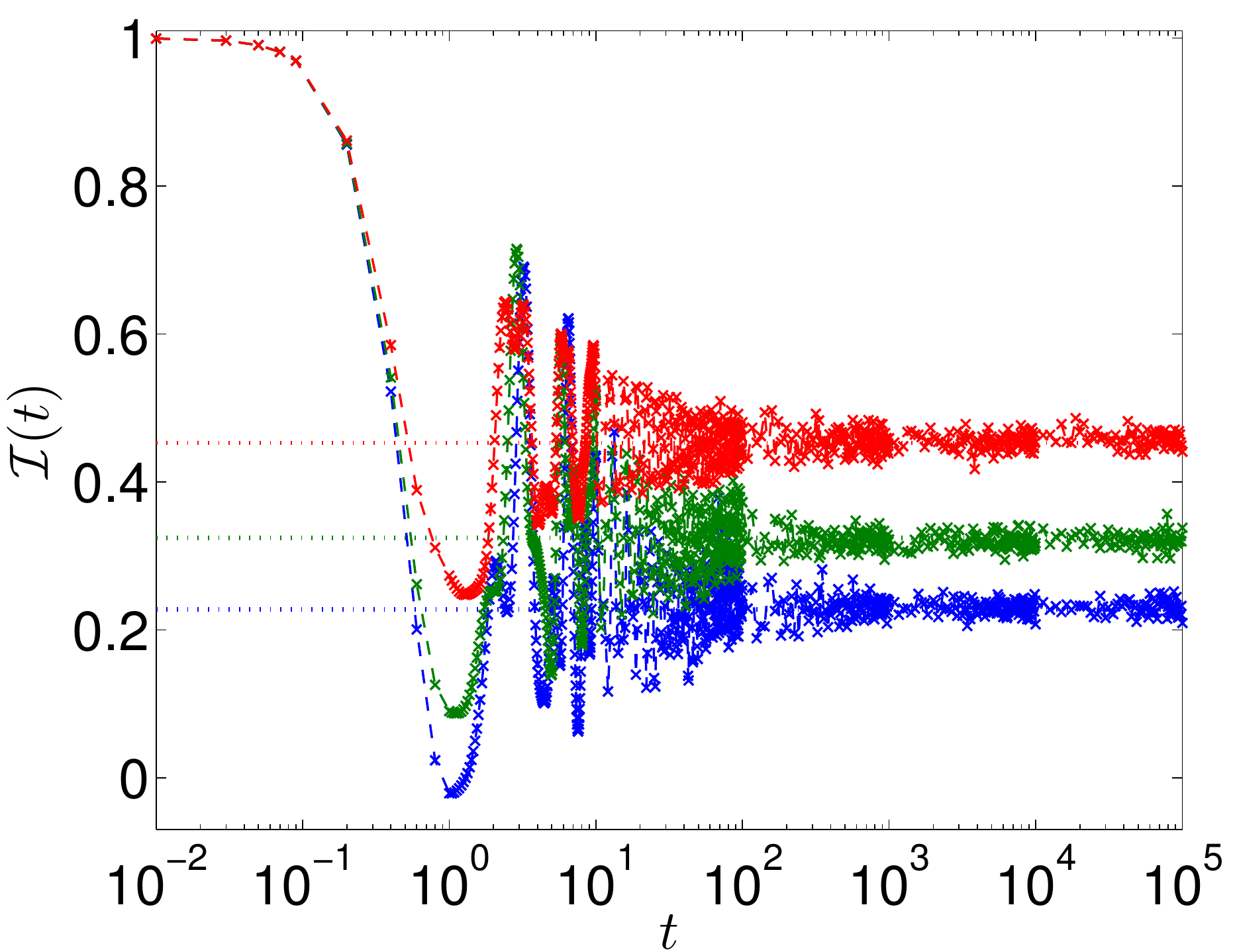} 
  \caption{Time evolution for the concurrence (top) and the imbalance (bottom) in the 
    AL phase (the values of $\Delta$ are chosen accordingly). 
    The system has $L=24$ sites, and results have been averaged over $10^2$ realisations 
    of the pseudo-disorder. The colour code of the two panels is the same.}
  \label{fig:non.interacting}
\end{figure}

We first consider the case with $V=0$. In the absence of interactions, the Hamiltonian 
reduces to a quadratic fermion model: its dynamics can be easily studied through 
the evaluation of the corresponding two-point correlation functions. 
In Fig.~\ref{fig:non.interacting}, the concurrence and the imbalance are plotted as a function 
of time for different disorder strengths. We only present the case $L=24$, in order to be consistent 
with the numerical simulations of the interacting systems: we simulated even larger lattices lengths 
without seeing appreciable differences (see Appendix~\ref{finitesize}).
After a non-trivial transient, that will be discussed later, 
both the concurrence and the imbalance saturate oscillating around a stationary condition 
that depends on the value of $\Delta$. A key observation is that both of them roughly saturate 
at the same time: in the Anderson insulator the entanglement does not evolve in time, 
in the regime where the spin dynamics is frozen. 
We will see that in the MBL phase the behaviour is qualitatively different.

\begin{figure}
  \includegraphics[width=8.2cm]{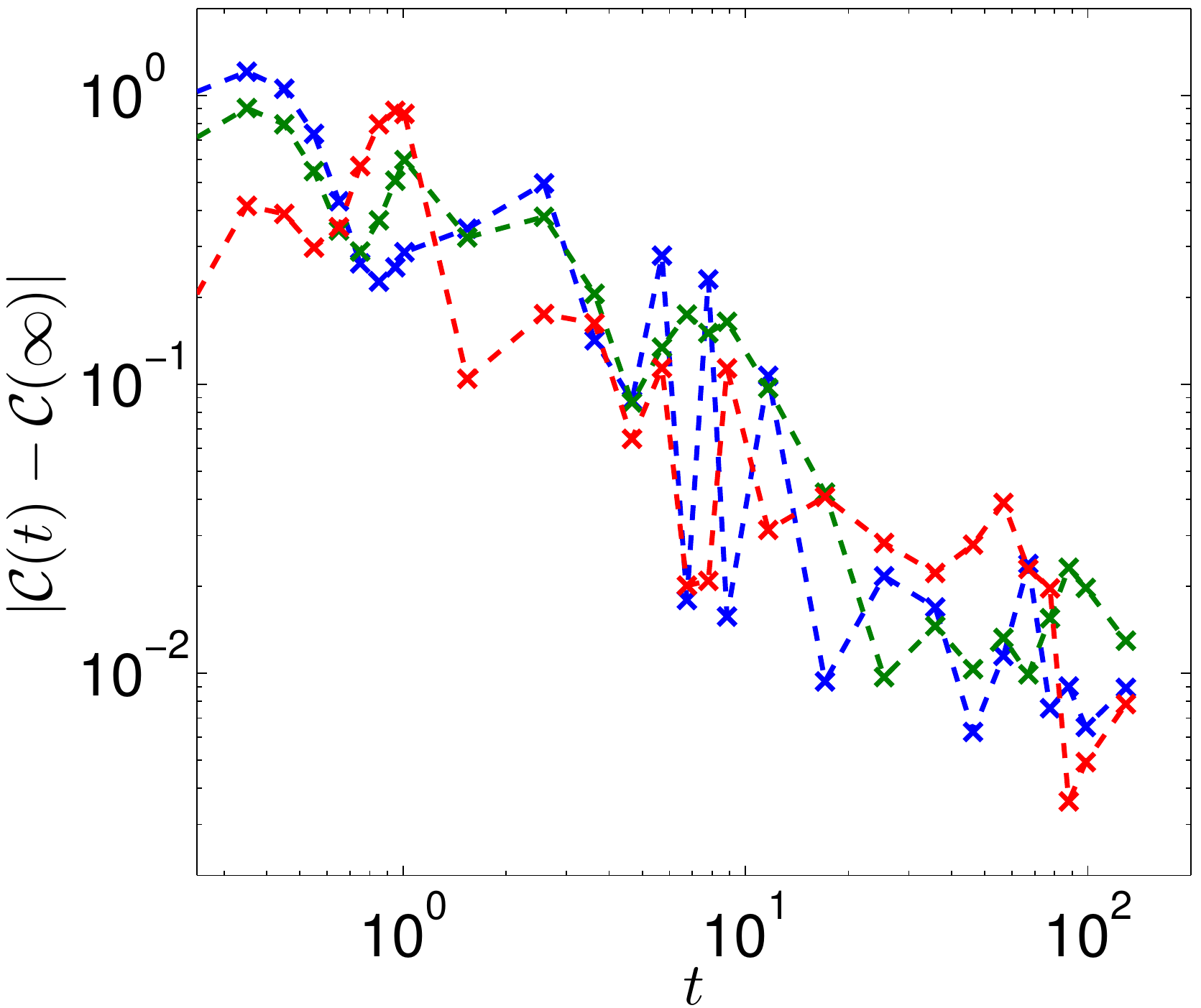} 
  \includegraphics[width=8.4cm]{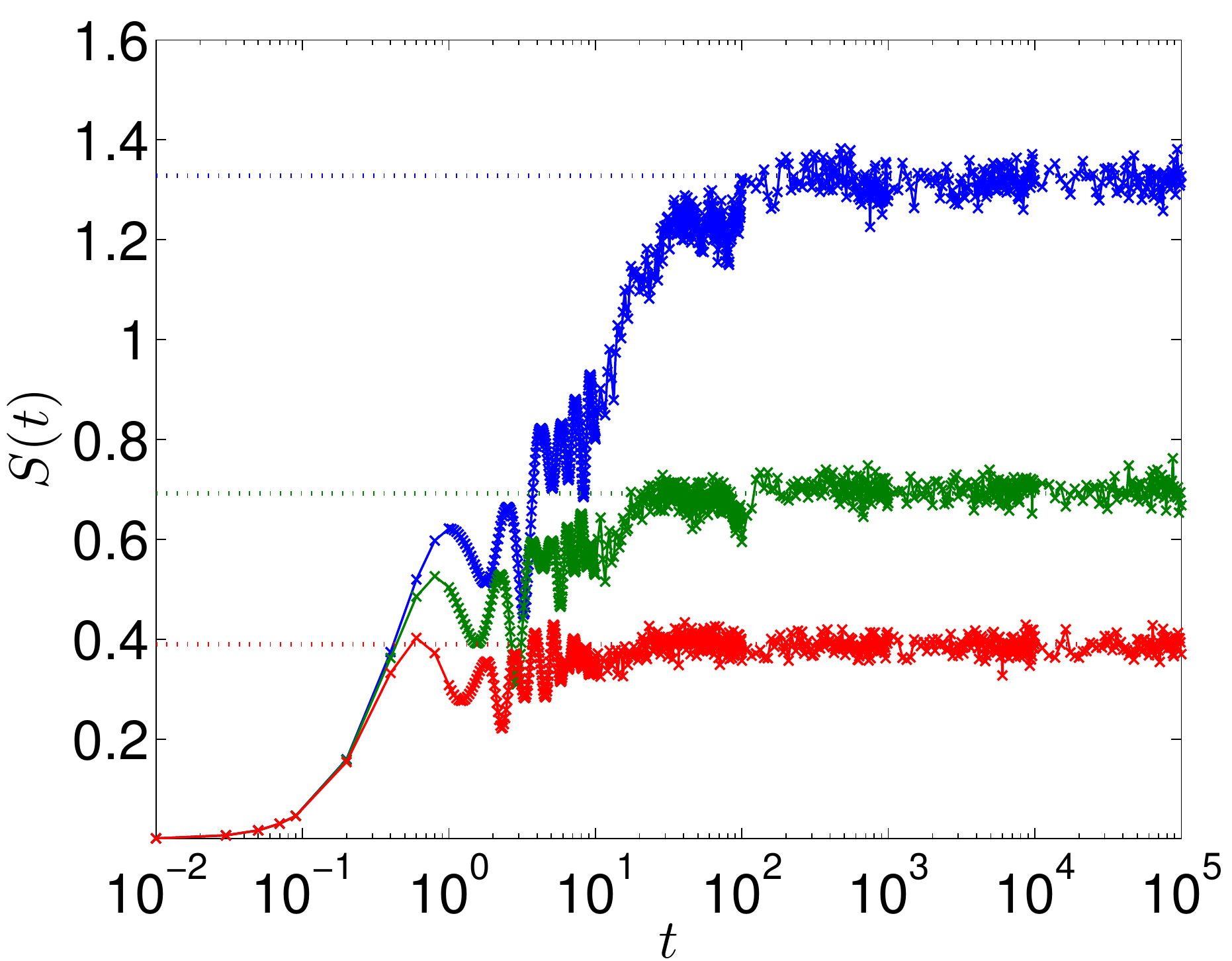} 
  \caption{(Top) Decay of the concurrence to its long-time stationary value for $t \gtrsim 1$. 
    In this regime, we observe a power law with an exponent independent of $\Delta$. 
    The concurrence has been averaged over small time bins in order to make the decay clearer.
    (Bottom) Half-chain entanglement entropy in the AL phase as a function of time. 
    In the regime in which the concurrence decays to its stationary value, the entanglement 
    entropy and the imbalance (Fig.~\ref{fig:non.interacting}, bottom panel) also evolve in time; 
    later everything saturates to a stationary value. 
    Numerical values and colour code are the same as in Fig.~\ref{fig:non.interacting}.}
  \label{fig:non.interacting-2}
\end{figure} 

In analogy with the stationary value of the imbalance~\cite{schreiber2015observation}, 
the corresponding two-site entanglement is also larger on increasing $\Delta$ and moving deeper 
in the localised phase. This behaviour can be qualitatively understood as follows. 
Starting from the factorised state of Eq.~\eqref{initial}, the short-time ($t \lesssim 1$) 
increase of entanglement is almost independent of $\Delta$ and is essentially due to the exchange 
coupling terms (the hopping in the fermion language) in Eq.~\eqref{hamiltonian}. 
Due to the many-body dynamics, the two-site entanglement then starts to decrease until a time $t^\star$, 
after which its subsequent dynamics is frozen. We find that $t^\star$ decreases with $\Delta$ 
(Fig.~\ref{fig:non.interacting}, top panel): the larger is $\Delta$ (that is, the deeper the system 
is in the localised phase), the earlier the concurrence will freeze, attaining indeed 
a larger stationary value. The time at which the dynamics is frozen, $t^\star$ 
should diverge when $\Delta\to 2$ as a power law, i.e. $t^\star \sim (\Delta-2)^{-\nu}$. 
We did not perform a detailed analysis as this aspect is tangential to the core of the work.

We have observed that, after the initial dynamics ($t\sim 1$), the concurrence exhibits a small decay 
until its saturation. For later comparison with the MBL case, it is useful to have a closer look 
at this intermediate regime. To this aim we plot both the two-site entanglement 
(Fig.~\ref{fig:non.interacting}, top panel) and the block entropy 
(Fig.~\ref{fig:non.interacting-2}, bottom panel). 
Moreover, to better analyse the decay of the concurrence, we subtract its long-time value 
(Fig.~\ref{fig:non.interacting-2}, top panel). In this intermediate regime, 
both the entropy and the concurrence, as well as the imbalance, evolve in time: 
the system has not yet frozen. On the opposite, we will see in the MBL phase that 
the two-site entanglement shows a power-law decay only in the long-time limit and not 
in the intermediate regime. Moreover, in this asymptotic regime, the spin dynamics is frozen 
and MBL dephasing takes place.
Therefore, the two power-law decays in the AL phase and in the MBL phase are different phenomena, 
which have to be distinguished from each other: the first one is a transient effect occurring 
before the entanglement and spin dynamics freeze, the second one is an asymptotic phenomenon 
occurring when the spin dynamics has already frozen.
Another clear difference with the MBL phase emerges also in the decay of the concurrence 
to its long-time limit: apparently the exponent of the power law does not depend 
on the disorder strength (Fig.~\ref{fig:non.interacting-2}, top panel); 
we are going to see that the power-law decay in the MBL phase behaves very differently.

\subsection{Many-body localised phase}
Equipped with the results for the AL phase, we are now going to discuss what happens 
in the presence of interactions. We first consider parameters for which the system 
is in the MBL phase (see the phase diagram in Appendix~\ref{diagram}). 
As outlined in the introduction, the spin dynamics in the MBL phase is frozen but correlations 
evolve in time due to many-body dephasing which is connected to the existence of an extensive number 
of local integrals of motion. 
Such dephasing peculiarly affects the behaviour of the entanglement: a signature of this phenomenon 
can be seen in the evolution of the half-system entanglement entropy which increases 
logarithmically in time only in the MBL phase. We are going to show that signatures 
of these effects can also be seen in the two-site entanglement, which shows a very special 
long-time behaviour unique to the MBL phase.

\begin{figure}[!t]
  \includegraphics[width=8cm]{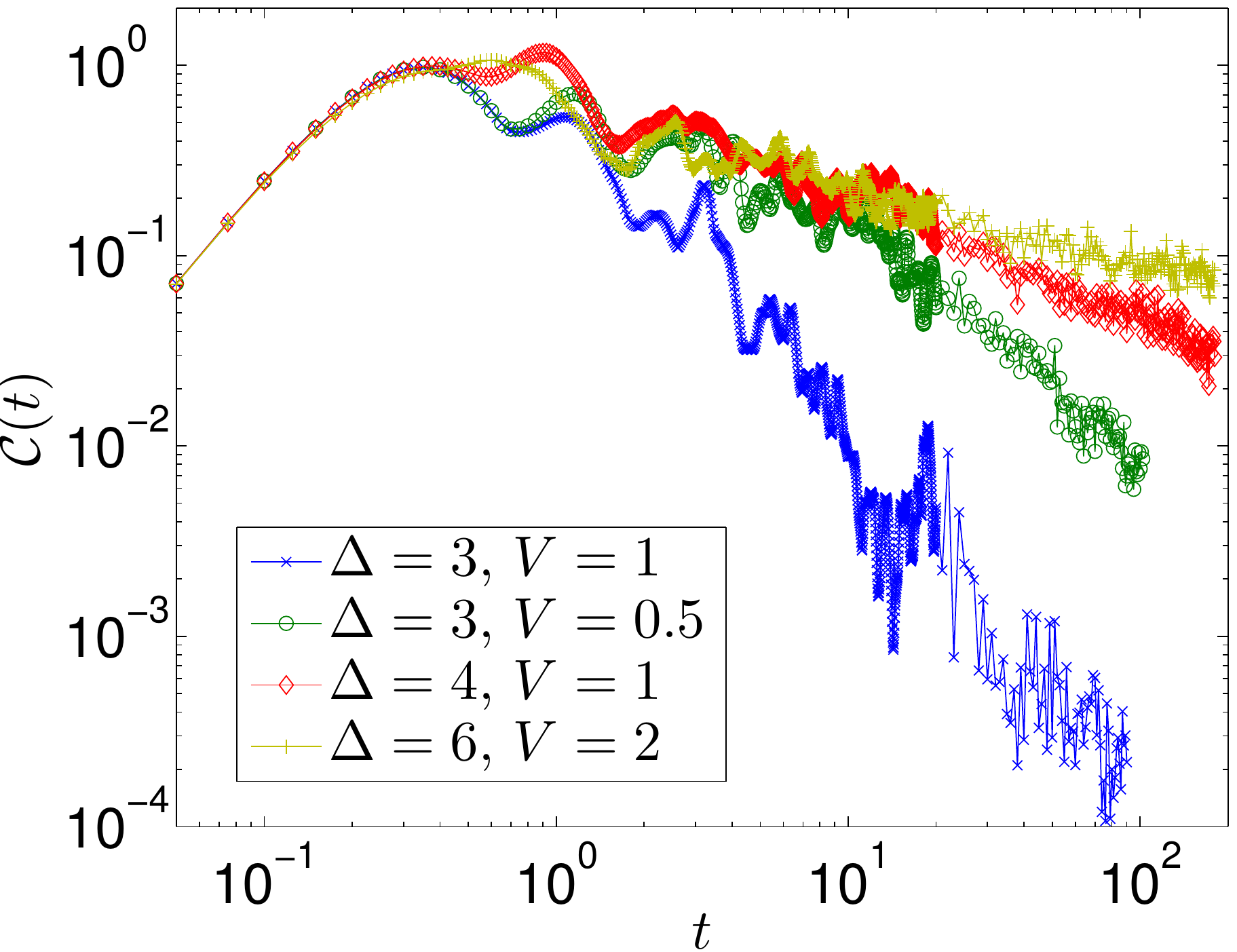}
  \includegraphics[width=8cm]{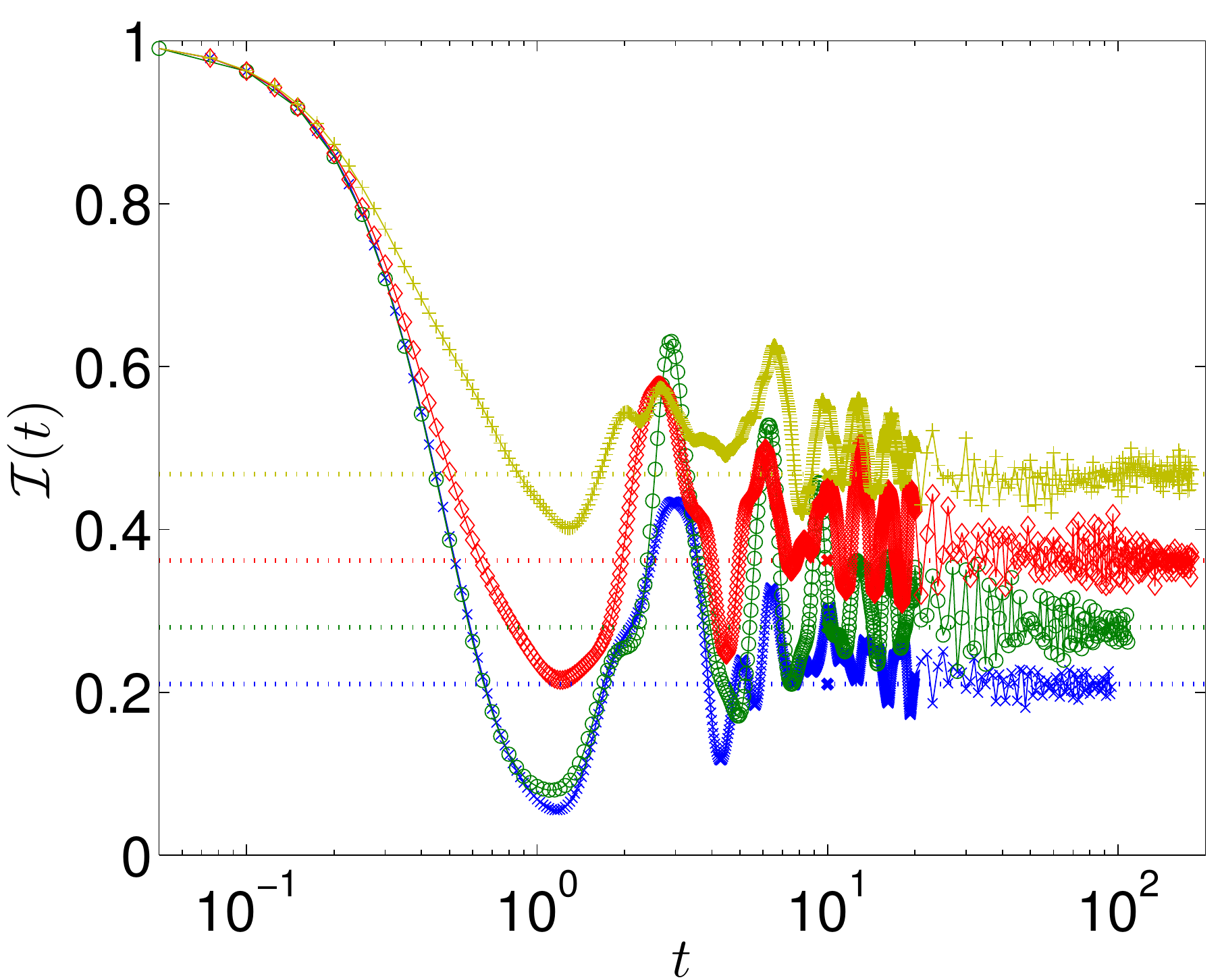}
  \caption{Time-evolution for the concurrence (top) and imbalance (bottom) in the MBL phase. 
    The plot for the concurrence is set in log-log scale in order to highlight the power-law decay. 
    The system has $L=24$ sites and different strengths of the pseudo-random potential 
    amplitude $\Delta$ are considered. 
    Data are averaged over $30$ realisations of pseudo-disorder.}
  \label{fig:conc.decay.all}
\end{figure}

In Fig.~\ref{fig:conc.decay.all}, we show the dynamical behaviour of the concurrence (top panel) 
and the imbalance (bottom panel) in the MBL phase. As previously discussed for the AL phase, 
also here the initial dynamics, up to $t \lesssim 1$, is independent of $\Delta$ and $V$: 
in this time regime, the only relevant terms of the Hamiltonian in Eq.~\eqref{hamiltonian} 
are those containing the exchange couplings.

The interesting regime occurs for longer times, $t \gtrsim 1$. In a region where the imbalance 
is already frozen, we clearly see a power-law decay for the concurrence with an exponent 
that depends both on $V$ and $\Delta$. This decay has to be contrasted with the saturation 
observed in the AL phase. Indeed we found that, in the presence of interactions, there is a regime 
where transport is absent but still the two-site entanglement evolves in time. 
It is important to stress that the dephasing mechanism which leads to the power-law decay 
of the two-site entanglement is the same as the one giving 
rise to the logarithmic growth of the entanglement entropy. 
We will discuss in detail this mechanism in Sec.~\ref{lbit-model}. 
The important new ingredient is that the two-site entanglement is ``easy'' to be measured. 
The comparison of Fig.~\ref{fig:non.interacting} and Fig.~\ref{fig:conc.decay.all} shows that, 
while the imbalance saturates both in the AL and MBL phases, the concurrence behaves 
qualitatively different in the two cases. 

The differences between AL and MBL can be further highlighted in the weakly 
interacting limit $V\ll 1$. In this case, two regimes appear in the dynamics of concurrence. 
After the common transient, the concurrence reaches a plateau typical of the AL phase. 
The plateau occurs for times $1 \le t \le t_{\rm int}\sim 1/V$ (for weak interactions 
it is possible to separate this time scale). Only at later times, $ t \ge t_{\rm int}\sim 1/V$, 
interactions set in and the concurrence starts to decay as a power law. 
A detailed analysis of this regime, together with the determination of $t_{\rm int}$ as a function 
of the interaction $V$, is presented in Appendix~\ref{weak}.

One final comment is in order. The results presented here are for the average 
two-site entanglement; its statistics is expected to be very interesting, as well.
From perturbative constructions of the integrals of motion~\cite{ros2015integrals},
it is known that the mechanism for delocalisation (and in general for mixing distant spins 
with a local spin) is to construct long non-local operators which, at a variance with respect 
to the AL case, are not simply bilinear in the raising and lowering operators.
Therefore the dynamics of one spin gets mixed with a line of spins of length $\xi_{\rm MBL}$, 
that is the MBL localisation length. In this way, the statistics of recurrences 
of the concurrence (the times at which $C_{i,j}$ returns to be non-zero) 
is less regular than in the AL case: after averaging, this leads to the power-law decay.

\subsection{Ergodic phase}
\begin{figure}[!t]
  \includegraphics[width=8cm]{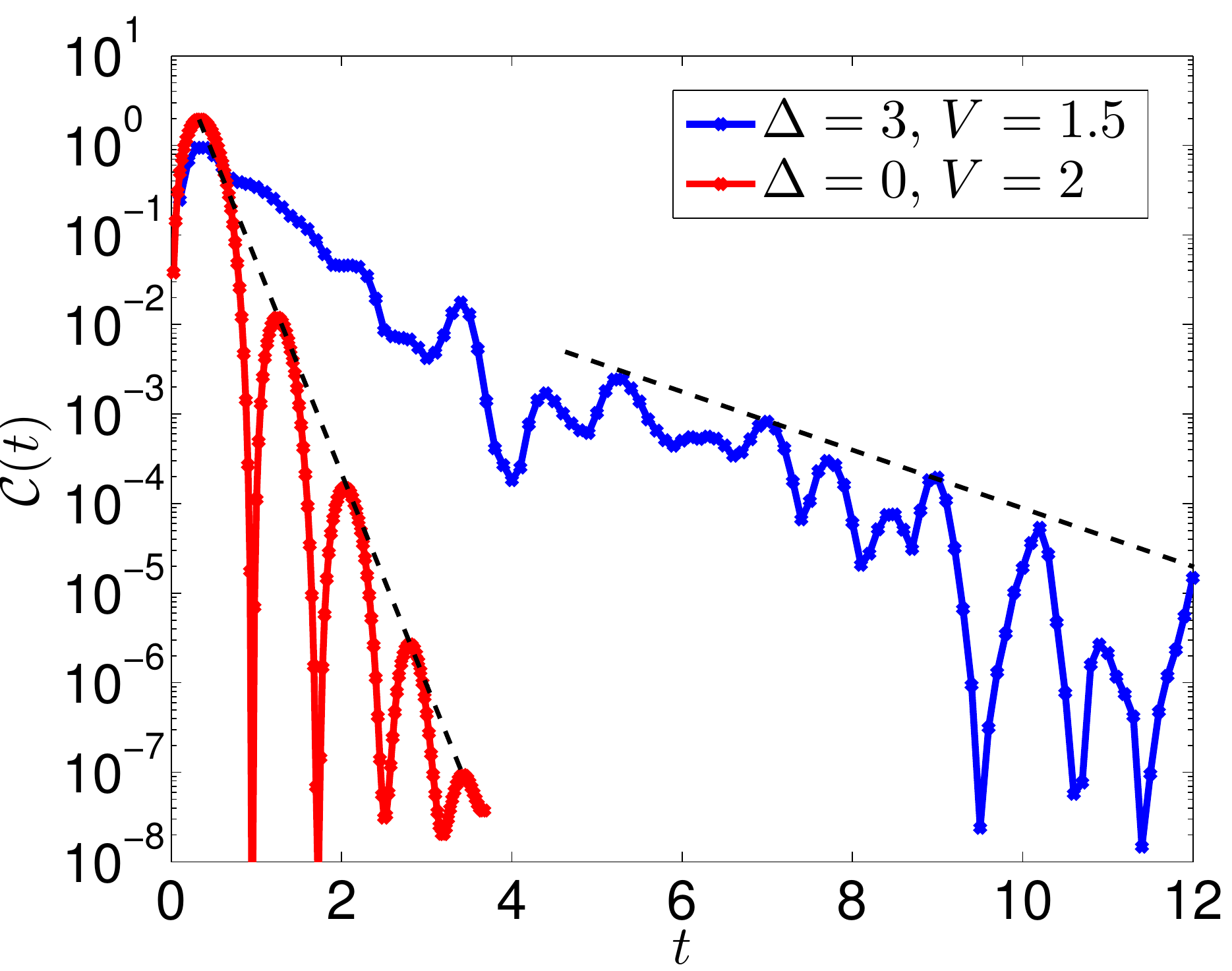}
  \includegraphics[width=8cm]{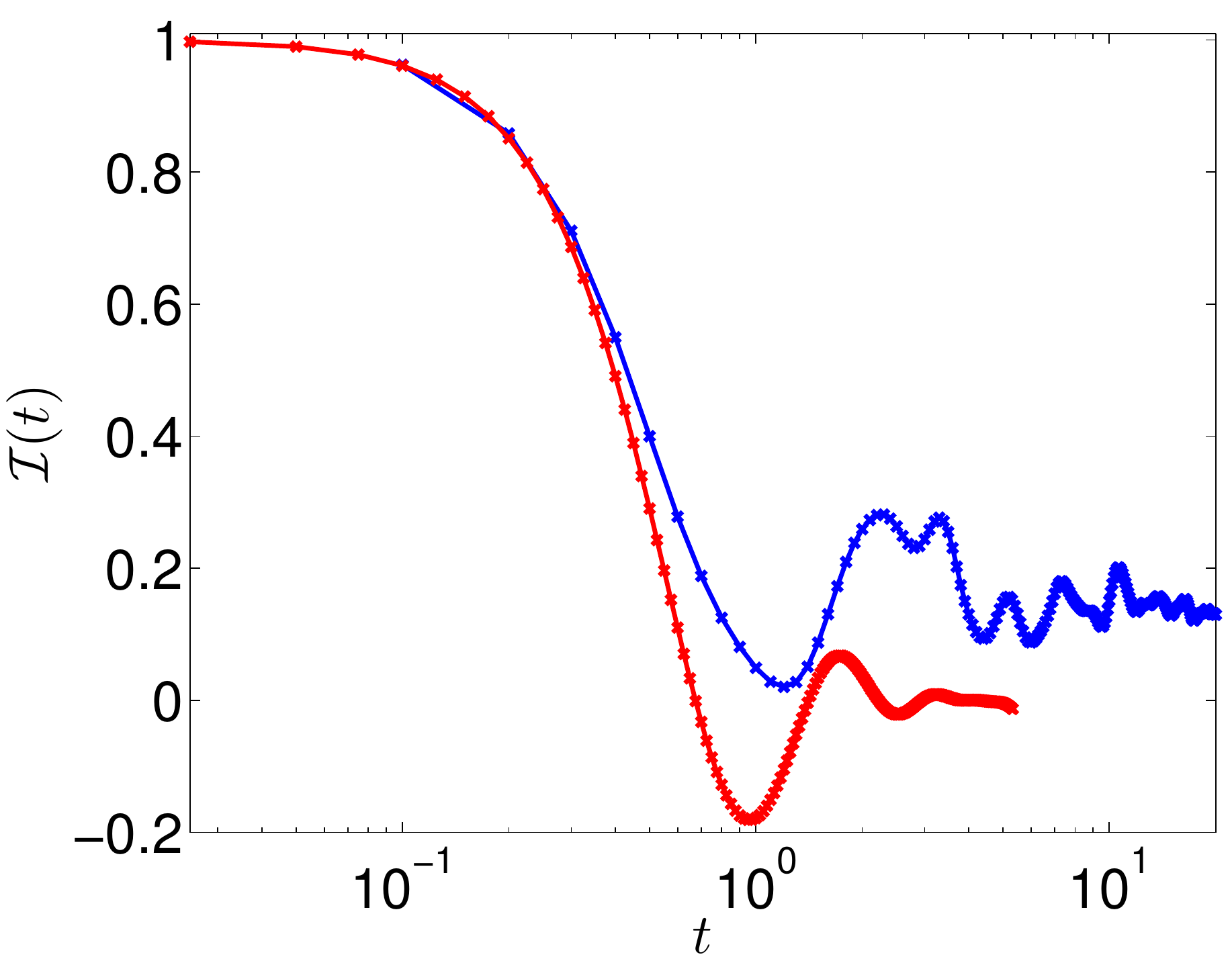} 
  \caption{Time-evolution for the concurrence (top) and the imbalance (bottom) in the ergodic phase. 
    The system has $L=24$ sites for $\Delta=3,\,V=1.5$, while it is slightly larger, having
    $L=48$ sites, for $\Delta=0,\,V=2$. Our results are averaged over
    $\sim 30$ realisations of the pseudo-disorder. 
    Black dashed lines in the top panel are a guide to the eye, to indicate the exponential decay.}
  \label{fig:L48N24.V2D0}
\end{figure}

We conclude this section by analysing the concurrence in the ergodic phase. 
In this case we consider interactions $V \gtrsim 1$. In this regime, TEBD simulations 
are more demanding, and we are able to follow the dynamics reliably only up to $t \sim 10$. 
In Fig.~\eqref{fig:L48N24.V2D0} (bottom panel) we show the imbalance as a function of time. 
In the ergodic phase it should go to zero in the long-time limit. As we can see,
times $t > 10$ are needed for a full equilibration. However the data clearly show that 
the imbalance is still decaying towards its stationary value. 

The behaviour of the concurrence (Fig.~\ref{fig:L48N24.V2D0}, top panel) qualitatively differs 
from the previous cases: here it vanishes abruptly. The data seem to indicate an exponential decay 
(especially visible for $\Delta = 0$). Due to the very fast decay, 
it is hard to unambiguously distinguish between an exponential and a high-order power-law decay. 
The ergodic phase is further characterised by large revivals with a typical period 
of the inverse of the exchange coupling.
In the long time (stationary) limit, the concurrence is expected to vanish. 
 The system will equilibrate to an effective temperature which is related
  to the energy initially injected in the system. 
This effective temperature is much larger than one (in units of $J$) 
for the choice of our initial states. At this temperature, any trace of thermal entanglement 
has already disappeared~\cite{amico2008entanglement}.

The different time-dependence in the decay of two-site entanglement is intimately connected 
to the monogamy of entanglement. Because of the much faster propagation of excitations 
in the ergodic phase, entanglement will spread faster. Consequently two-site entanglement 
will decay rapidly, that is, exponentially as observed in the simulations. 
The spread, and related decay in the concurrence, is slower in the MBL phase.

\section{Concurrence in the $\ell$-bit model}
\label{lbit-model}
In order to corroborate our numerical results, we are going to show that the above discussed 
phenomenology can be obtained by means of an effective model expressed in terms of the local 
integrals of motion. Within this effective model it is possible to obtain semi-analytical
results and, most importantly, it is natural to link the behaviour of the two-site entanglement 
to the dephasing that is typical of the MBL phase. 

As already mentioned in the introduction, a key feature of the MBL phase is that it possesses 
an extensive number of local integrals of motion. This notion of integrability leads to 
a very insightful description of the system in terms of an effective 
phenomenological $\ell$-bit model~\cite{Abanin.2103.LIOM,Nandkishore:2015aa}:
\begin{equation}
  \hat H = \sum_j h_j \hat \tau_j^z + \sum_{j \neq l} \mathcal{J}_{j l} \hat \tau_j^z \hat \tau_l^z + \ldots \,.
  \label{ell.bit.model}
\end{equation}
Here, $\{\hat \tau_j^x, \hat \tau_j^y, \hat \tau_j^z\}$ are the localised spin-$1/2$ operators 
associated to the local integrals of motions (in this context, they are also called 
the $\ell$-bit operators: ``$\ell$'' stands for localised). In the previous definition, $
h_j$ are random fields, and $\mathcal{J}_{j l} = \mathcal{W}_{j l} \, e^{-\alpha |j-l|}$
are the interaction terms, with $\mathcal{W}_{j l}$ assumed to be a random variable 
uniformly distributed in the range $[-W,W]$~\cite{Note1}.
Further terms in the Hamiltonian include $n$-body interactions, with $n>2$, which, 
for simplicity, we will not consider here and are irrelevant for our purposes (see later). 
The spins $\hat{\boldsymbol{\tau}}_i$ are local functions of the physical spins: 
the precise form of this mapping is not important for the present work. 
In an Anderson insulator, the couplings $\mathcal{J}_{j l}$ are vanishing, and a set of 
independent non-interacting spins is sufficient for an effective description of the dynamics 
on a distance larger than the localisation length. 

The analysis of the model in Eq.~\eqref{ell.bit.model} gives us the possibility to see 
from a different perspective, and to clearly understand, the difference in the behaviour 
of the two-site entanglement between AL and MBL phases.
From one side, the Hamiltonian of an Anderson insulator will only lead to single-bit rotations 
(that do not modify the entanglement). On the opposite, in the MBL phase, the second term 
of Eq.~\eqref{ell.bit.model} is responsible for two-qubit gates (controlled phase-shifts) 
that lead to a time-dependence of the entanglement. These terms are the ones leading 
to the logarithmic growth of entropy~\cite{znidaric2008many,Bardason2012,Abanin.2013.EntGrowth} 
(higher-order contributions to the Hamiltonian do not change the picture). 
We are going to show that they also lead to the power-law decay of the concurrence. 

The dynamics generated by the phenomenological $\ell$-bit model gives a meaningful comparison 
with the exact dynamics of Hamiltonian~\eqref{hamiltonian} for times $t > 1$. 
In this time-regime, the interactions and the quasi-periodic Aubry-Andr\'e potential become 
relevant (in the initial transient we saw that only the exchange terms affect the concurrence dynamics).

\begin{figure}[!b]
  \includegraphics[width=8cm]{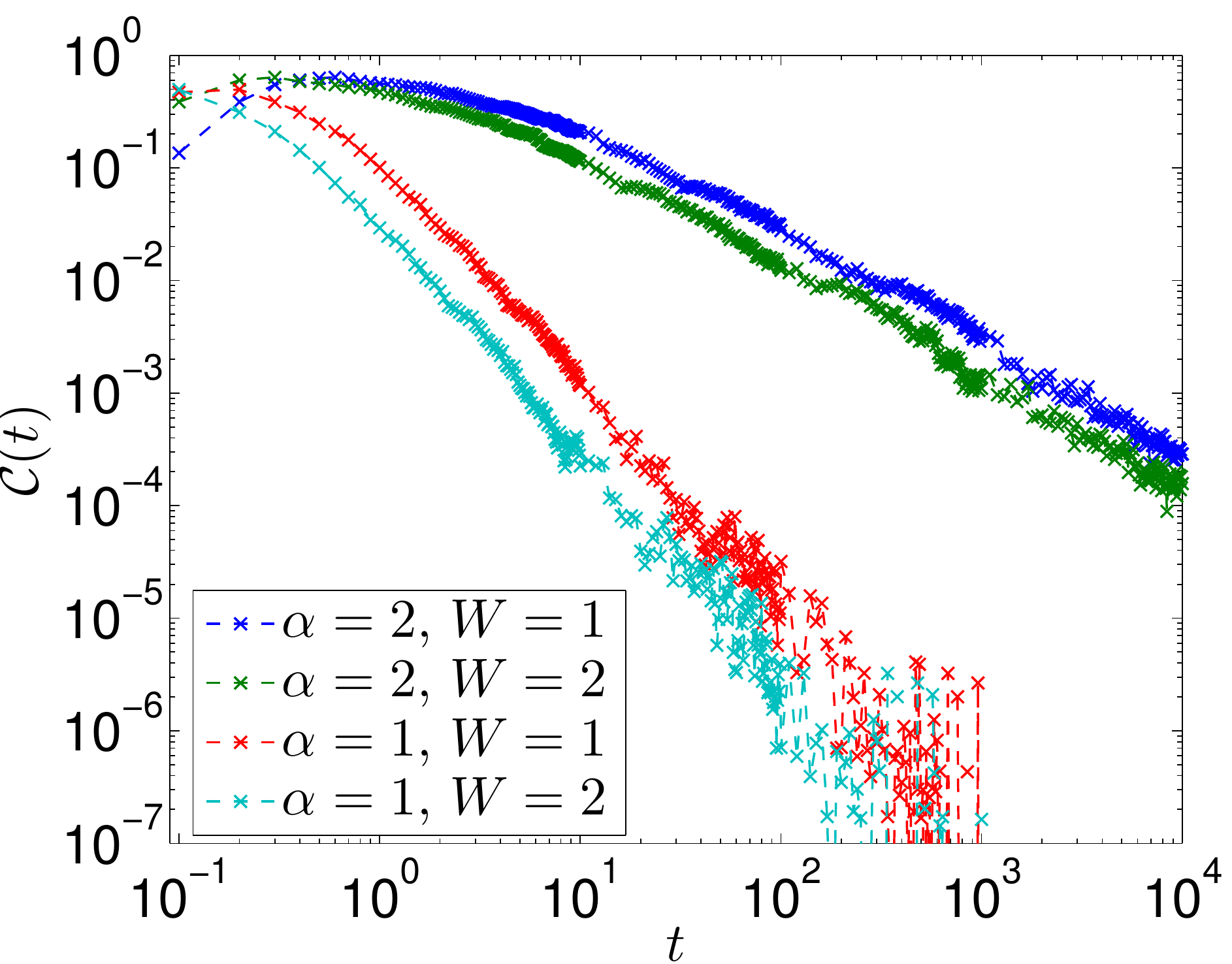}
  \caption{Time-evolution for the concurrence in the $\ell$-bit model~\eqref{ell.bit.model}. 
    The log-log scale highlights the power-law decay. We consider a system with $L=72$ sites, 
    and average over $10^2$ simulations where the initial state as well as 
    the couplings in the Hamiltonian~\eqref{ell.bit.model} are chosen randomly.}
  \label{fig:l.bit.concurrence}
\end{figure}

The dynamical protocol we consider goes as follows. The system is initially prepared in a generic 
separable state given by
$$
 \ket{\psi_0} = \otimes_{j=1}^L \left[ \cos(\phi_j) \ket{\uparrow} + e^{i\theta_j}\sin(\phi_j)\ket{\downarrow} \right] 
$$
where $\{\ket{\uparrow},\ket{\downarrow}\}$ are the eigenstates of the $\hat \tau_j^z$ operator 
and $\phi_j$, $\theta_j$ are randomly chosen parameters. 
The time evolution of the $\ell$-bit operators can be easily computed 
(see Appendix~\ref{corr-lbit}) and the concurrence can be determined as a function of time. 
In our analysis, we average over distinct initial states (different realisations of 
$\{\theta_j\}$, $\{\phi_j\}$), local disorder terms ``$h_j$'' (despite it has 
absolutely no effect on the concurrence) and interacting terms ``$\mathcal{W}_{j l}$''. 

As already mentioned, when we are in the AL phase, the Hamiltonian~\eqref{ell.bit.model} 
induces a local dynamics: it cannot lead to any change in the entanglement.
For the MBL phase the situation is much more intriguing, because of the coupling between 
the $\ell$-bit operators. Figure~\ref{fig:l.bit.concurrence} displays the concurrence 
as a function of time, averaged over random realisations of the external fields, couplings 
and initial preparation of the state. We see that the concurrence decays as a power law, 
fully confirming the fact that this form of two-site entanglement behaviour is a typical feature of MBL.

\section{Experimental issues and a bound for the concurrence}
\label{bound}
Experimentally, the detection of concurrence for unknown two-site reduced density matrices 
might face some imperfections, such as the lack of complete experimental control in the measurements, 
on-site number fluctuations or thermal smearing. For a better comparison with the experiments,
it is important to consider all these issues. 

The spin-model of Eq.~\eqref{hamiltonian} does not include number fluctuations, 
that are present in the native two-species Bose-Hubbard Hamiltonian. 
A detailed analysis of these effects has been performed 
in Ref.~[\onlinecite{Mazza_NJP}]. As long as the on-site repulsion between bosons is much larger 
than their hopping (in practice, a factor five in this ratio is enough) the predictions 
of the effective spin models are reliable. As far as thermal fluctuations are concerned, 
the analysis of Ref.~[\onlinecite{Mazza_NJP}] confirms that, as expected, our model is reliable 
for experiments if the temperature is of the order of few percents of the on-site interaction. 

Here it is of particular importance to address the lack of control in the pulses that are
needed to measure the entanglement: this fact leads to the detection of smaller correlation values. 
In connection with this issue, below we provide a very useful lower bound for the concurrence. 
For the present model, the qualitative behaviour of this bound agrees with the actual concurrence 
dynamics with high fidelity. For longer times the agreement becomes even quantitative, 
since the bound becomes tighter with increasing time.

\begin{figure}[!t]
  \includegraphics[width=8.1cm]{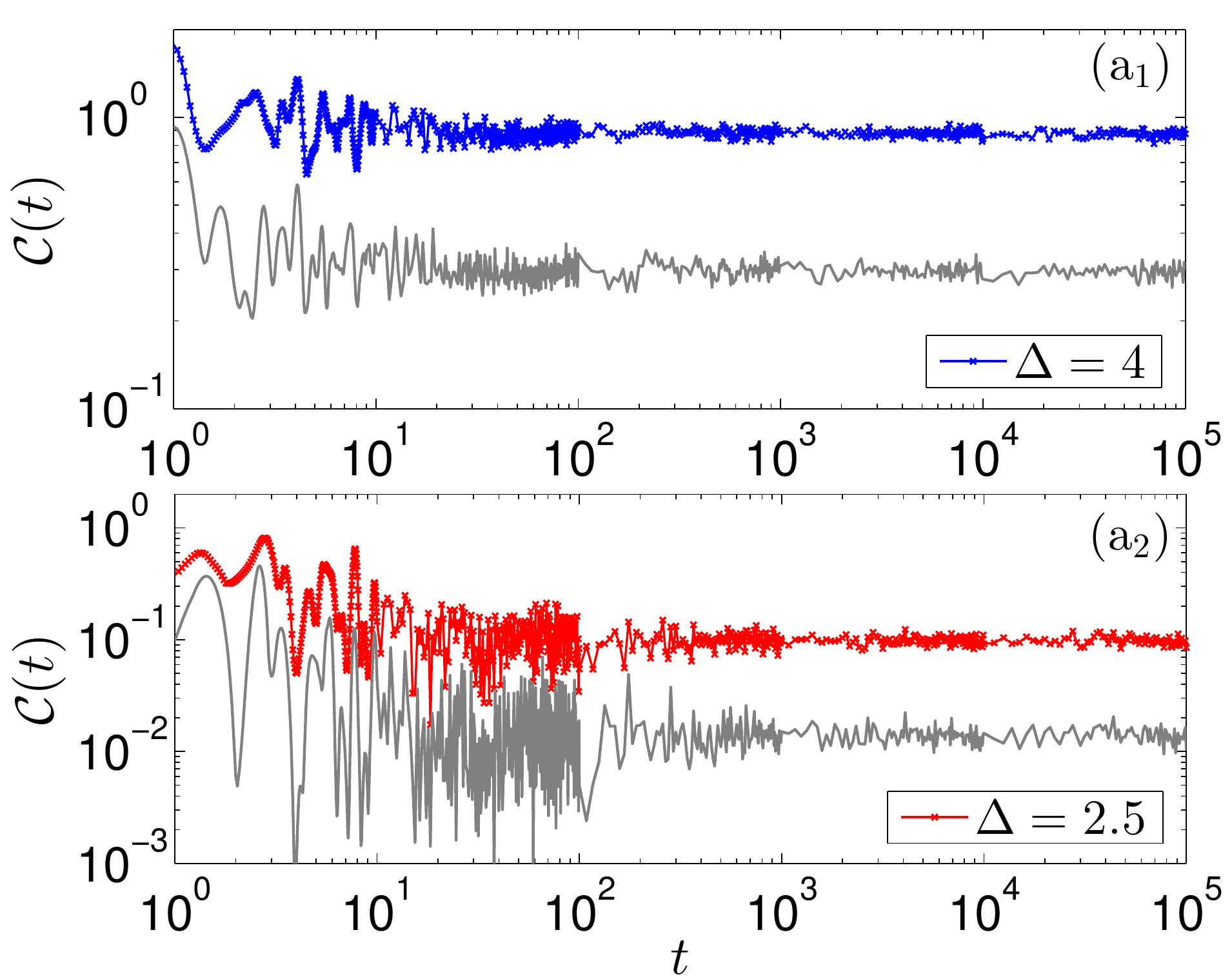}
  \includegraphics[width=7.7cm]{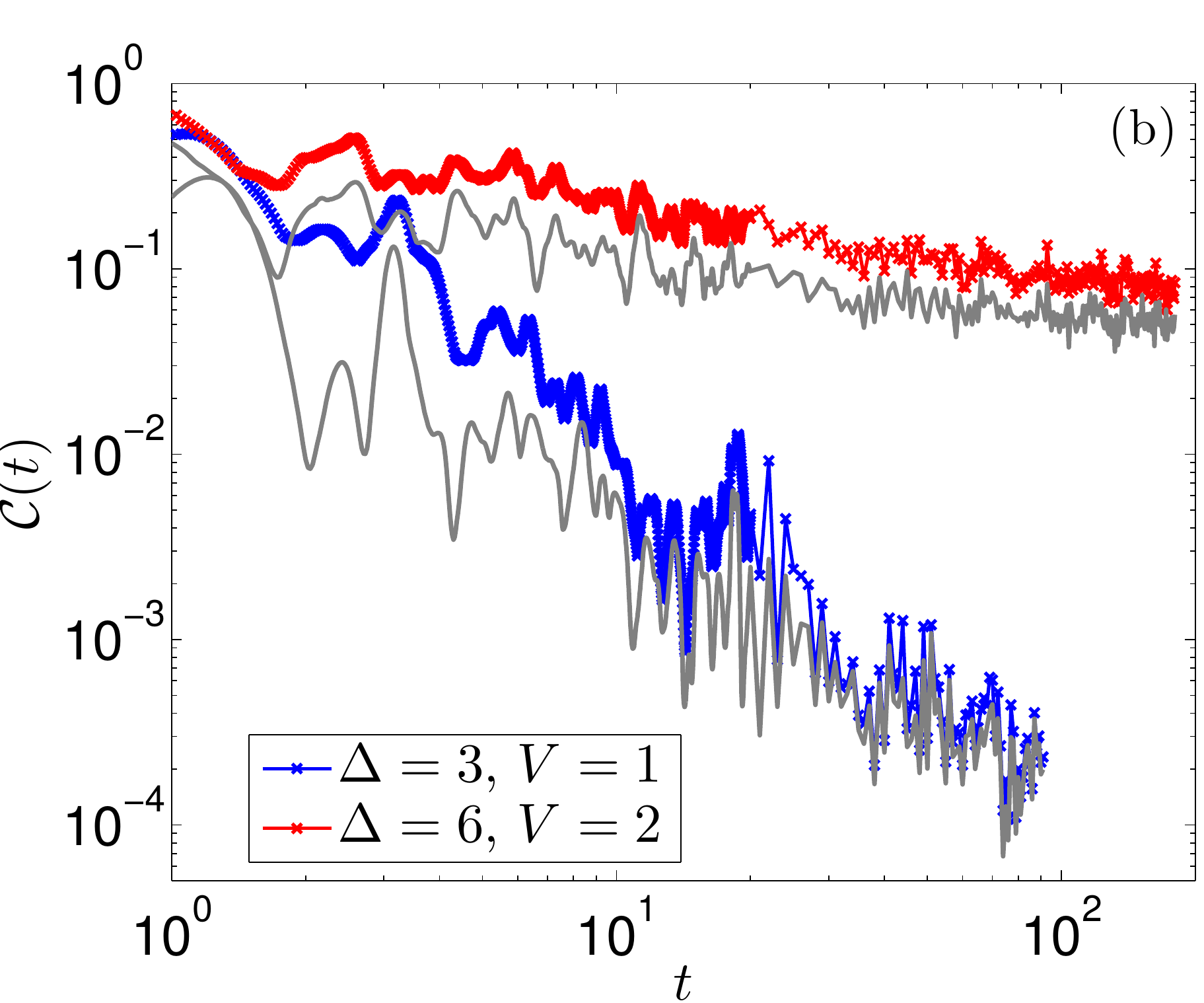}
  \includegraphics[width=8.1cm]{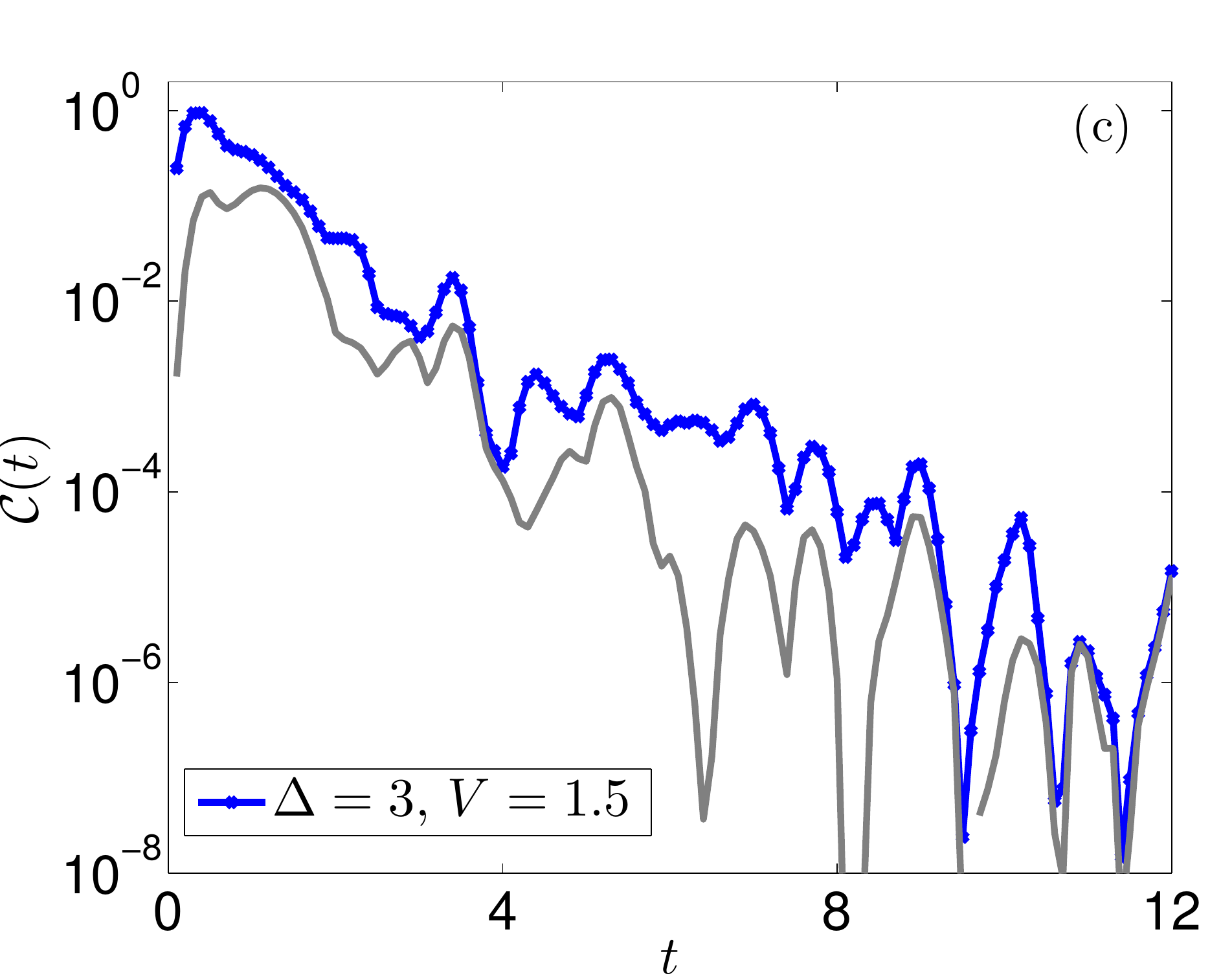}
  \caption{Time evolution for the concurrence (colour) and its experimental bound (grey) 
    based on measurements performed with global pulses on both sites (see main text), 
    in the three distinct phases: AL (panels a$_1$ and a$_2$),
    MBL (panel b), and ergodic (panel c).}
  \label{concurrence-bound}
\end{figure}

A full two-site reduced density matrix can be obtained by measuring all its spin-spin 
correlation functions. In principle, such measurements could be performed in a cold-atom setup 
by first applying pulses on each individual spin site and then allowing it to freely rotate. 
After the appropriate time-interval has elapsed, the measurement is performed in a fixed 
basis (e.g., in the $z$-eigenstate basis $\{\ket{\uparrow},\ket{\downarrow}\}$). 
However, in cold-atom implementations of Hubbard models, pulses on individual spin sites 
are not yet implemented. In this case, only a global pulse on both spin sites is allowed, 
and consequently only measurements of spin-spin correlations along the same direction are performed. 
In this case, the reduced density matrix element $\rho_{\uparrow \downarrow}$ is approximated by
 \begin{equation}
   \widetilde{\rho}_{\uparrow \downarrow} = \langle \hat S_i^x \hat S_j^x + \hat S_i^y \hat S_j^y 
   \rangle = \Re (\rho_{\uparrow \downarrow}).
   \label{eq:rho.jl.bound}
 \end{equation}
Recalling the expression for the concurrence~\eqref{conc0:eqn}, 
since $|\widetilde{\rho}_{\uparrow \downarrow} | \leq | \rho_{\uparrow \downarrow} |$, 
we obtain the lower bound
\begin{equation}
  \widetilde{C}_{i,j} \equiv 2 \max\left[ 0,|\widetilde{\rho}_{\uparrow \downarrow}| 
    - \sqrt{P_{\uparrow \uparrow} P_{\downarrow \downarrow} }\right] \leq C_{i,j}\,.
\end{equation}
Since the concurrence is generated essentially only between nearest-neighbour sites, 
let us focus on this case. Here, $\rho_{\uparrow \downarrow}$ has a particularly simple form, 
$\rho_{\uparrow \downarrow} = \langle \hat a^{\dagger}_j \hat a_{j+1} \rangle$, 
and a clear physical interpretation. As we can see in Eq.~\eqref{eq:rho.jl.bound}, 
the bound $\widetilde{C}_{i,j}$ for the concurrence does not involve 
the imaginary parts of the hopping terms, 
which physically correspond to the spin current between the neighbouring sites 
$\propto \Im( \langle \hat a^{\dagger}_j \hat a_{j+1} \rangle)$. Since in 
the localised phase we should asymptotically expect no current (despite still having a flow 
of information/correlations), the above bound should become tighter with the increase of time. 
Precisely, $\widetilde{\rho}_{\uparrow \downarrow} \sim \rho_{\uparrow \downarrow} $ for $t \gg 1$, 
and consequently $\widetilde{C}_{i,j} \sim C_{i,j} $.

Figure~\ref{concurrence-bound} compares the dynamics of the concurrence (in colour) 
with its bound (grey) for the three different phases: AL (panels a$_1$ and a$_2$), MBL (panel b), 
and ergodic (panel c). It is evident that, except for the initial transient, 
the bound faithfully reproduces the behaviour of the two-site entanglement, 
thus making the experimental detection easier.

\section{Conclusions}
\label{conclusions}
The aim of this paper was to show that distinct features of the many-body localised phase 
can be detected through a measure of two-site entanglement. The time-dependence of the concurrence, 
the measure we used to quantify two-site entanglement, can clearly distinguish between MBL, 
AL and ergodic phases. In order to highlight this different behaviour, we studied the dynamics 
of two-site entanglement in a quantum quench, as experimentally implemented 
by Schreiber {\it et al.} in Ref.~[\onlinecite{schreiber2015observation}]. 
We stress the importance on the choice of the quantum correlation quantifier 
 as well as in the initial state, 
since different choices could lead to distinct long-time behaviours~\cite{Campbell.2016}.

Here we considered a two-species Bose-Hubbard model in an optical lattice, 
undergoing a quasi-periodic Aubry-Andr\'e potential. Ignoring number fluctuations, 
this system reduces to the XXZ model, studied here. Our results were based on 
time-dependent density matrix renormalisation group simulations complemented 
by semi-analytical calculations on an effective model.
After an initial transient, dominated by the kinetic term in the Hamiltonian, the concurrence dynamics 
in the different phases shows a strikingly different behaviour. The two-site entanglement 
saturates to a non-vanishing constant in the AL phase, while it decays as a power law 
in the MBL phase and exponentially fast in the ergodic one.
 
In order to corroborate the claim that the power-law decay is a characteristic trait 
of the MBL phase, we analysed the same problem using an integrable phenomenological $\ell$-bit 
model~\cite{Abanin.2103.LIOM,Nandkishore:2015aa}, which is known to capture the essence 
of the MBL phase. This is a consequence of the unitary equivalence
of any MBL system to an integrable one with localised integrals of motion. 
Exploiting the integrability of the phenomenological model, we could compute 
the two-site entanglement in a semi-analytical way, 
highlighting the same power-law decay occurring in the MBL phase of our system. 

The main advantage of our proposal relies on the fact that experimental protocols 
to measure the two-site concurrence have been already implemented, thus our analysis 
can be tested in the laboratory.
In order to make a closer contact with the experiment, we also computed a useful bound 
for the concurrence that can be more easily measured. In the relevant time-regime, 
this bound turns out to be very close to the actual value of the entanglement.

It is important to stress that all the results obtained in this work hold for averaged quantities. 
Single disorder realisations have very different aspects and the power-law decay 
itself of $\mathcal{C}$ comes from single realisations of disorder in which $C_{i,j}$ is mostly zero, 
except for occasional ``revivals''. 
It would be of great interest to analyse the full statistics of entanglement, 
something that is also experimentally accessible. 
In fact, it is possible that multifractal properties of the eigenstates 
in the MBL phase~\cite{buccheri2011structure,de2013ergodicity,luitz2016anomalous} 
are reflected in full counting statistics of these revivals and higher moments of $C_{i,j}$.

A further perspective of future work would be to understand the power-law behaviour 
when the transition to the ergodic phase is approached. Other interesting questions 
concern the behaviour of the concurrence when a local quench is performed, 
especially in connection with the phenomenon of the logarithmic light-cone propagation 
of correlations~\cite{Deng_arX16}.

\textit{Note added.} After the completion of this manuscript, we became aware of a related 
work~\cite{Tomasi2016} discussing the dynamics of the two-site quantum mutual information 
in the MBL phase.

\acknowledgments

We warmly thank Leonardo Mazza for useful discussions. This research was supported in part 
by the Italian MIUR via FIRB project RBFR12NLNA, by the EU integrated projects SIQS and QUIC 
and by ``Progetti interni - SNS''. F. I. acknowledges financial support from the Brazilian 
agencies FAPEMIG, CNPq, and INCT- IQ (National Institute of Science and Technology for Quantum Information).
 We acknowledge
the CINECA award under the ISCRA initiative, for the availability 
of high performance computing resources and support.

\appendix

\section{Phase diagram}
\label{diagram}
Depending on the value of its coupling constants, the model studied in this 
work [Eq.~\eqref{fermion.model}] presents three distinct phases: ergodic, MBL, and AL phases. 
Although several exact results~\cite{Mastropietro_PRL} on the ground state
and the phase diagram along the non-interacting line~\cite{Aubry-Andre} are known, 
the location of the phase boundaries has not been worked out so far. 
A detailed analysis of the phase boundaries and of their properties lies beyond the purpose 
of the present work. Here we only need a reliable analysis that enables us to unambiguosly choose 
the couplings in order to be in one of the three phases. 
Therefore, our aim is an (approximate) phase diagram for the Hamiltonian~\eqref{fermion.model}. 
We obtain it by combining different approaches. More precisely, we study: 
{\em i)} the time-dependence of the entanglement block-entropy for a bipartition of half 
the system size~\cite{znidaric2008many,Bardason2012,Abanin.2013.EntGrowth}, and 
{\em ii)} the level-spacing statistics (LSS) of the Hamiltonian~\cite{oganesyan2007localization}. 
A detailed discussion about the way these quantities can discriminate 
between the different phases can be found in the cited references. 

The entanglement entropy $S_{A}(t)$ of a block with $A$ sites is defined as
\begin{equation}
  S_{A} = - {\rm Tr}[\rho_{A} \log(\rho_{A})]\,,
  \label{l.block.ent}
\end{equation}
where $\rho_{A} = {\rm Tr}_{\neq A}[\rho]$ is the reduced state for the corresponding block sites. 
The increase in time of $S_A(t)$ behaves differently in the three phases. 
We expect a ballistic growth in the ergodic phase, in contrast to a logarithmic dependence in the 
MBL phase and a saturation in the AL phase~\cite{znidaric2008many,Bardason2012,Abanin.2013.EntGrowth}.

\begin{figure}[!t]
  \includegraphics[scale=0.45]{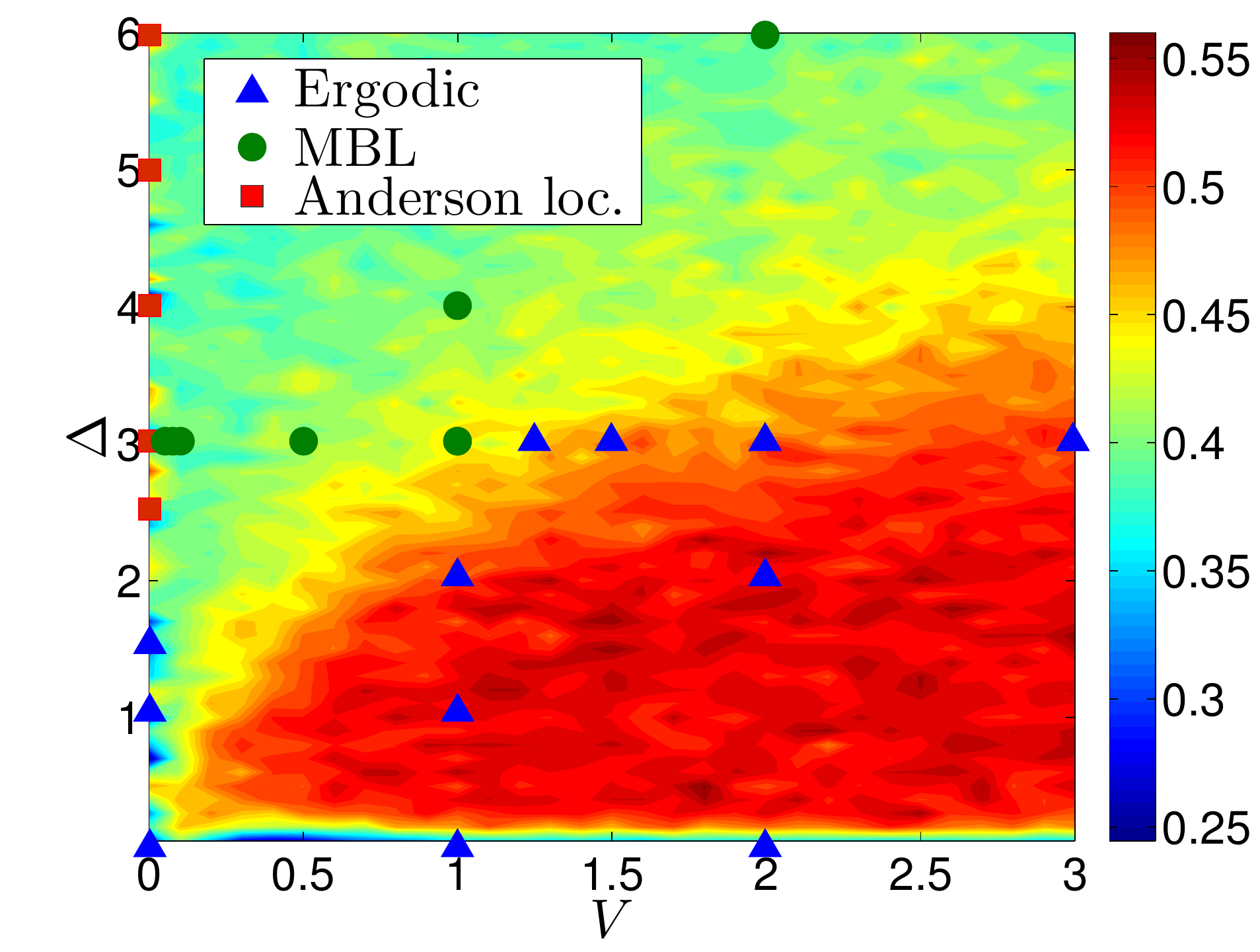}
  \caption{Phase diagram for the fermionic model~\eqref{fermion.model} at half filling 
    [corresponding to the spin model~\eqref{hamiltonian} at total spin $S^z_{\rm tot} = 0$]. 
    Marked points have been studied using TEBD simulations, for systems up to $L=24$ sites, 
    where the half-system entanglement entropy dynamics has been also analysed. 
    The colour filling corresponds to the average level statistics 
    $\langle r_n \rangle$ [see Eq.~\eqref{eq:rn_def}] for a system with $L=12$ sites. 
    Only in correspondence to the marked points we can reliably say that the system 
    is in one of the three different phases.}
  \label{appendix.fig:phase.diagram}
\end{figure} 

The rationale behind the spectral statistics approach~\cite{oganesyan2007localization} lies on the fact that 
the LSS follows a Wigner-Dyson distribution in an ergodic system, and a Poisson law 
in an integrable system. As we have extensively discussed, MBL is a special case 
of integrable system (the reader can find more details on the spectral statistics 
in Refs.~[\onlinecite{Haake:book,Berry_PRS76,Berry_LH84,Poilblanc_EPL93}]). 
In order to distinguish the various phases, instead of considering the whole LSS, we can restrict 
to a quantity whose average takes markedly different values on the two distributions. 
Having defined the gaps between adjacent many-body levels $\{E_n\}$ 
as $\delta_n = E_{n+1} - E_n \geq 0$, we define the ratio
\begin{equation}
  0 \leq r_n = \min\{\delta_n,\delta_{n+1}\}/\max\{\delta_n,\delta_{n+1}\} \leq 1\,.
  \label{eq:rn_def}
\end{equation}
The different phases are characterised by a different value of the average 
$\langle r_n \rangle$ over the level spacing distribution.
From the results of Ref.~[\onlinecite{oganesyan2007localization}], 
we expect to have $\langle r_n \rangle\simeq 0.386$ in the localised phase 
(Poissonian LSS distribution), and $\langle r_n \rangle\simeq 0.5295$ 
in the ergodic phase (Wigner-Dyson LSS distribution).

The colour code in Fig.~\ref{appendix.fig:phase.diagram} shows how this analysis 
can discriminate between the ergodic and the localised phase. 
In addition, for the points of the phase diagram marked by a symbol, 
we also studied the time-dependence of the entanglement entropy. 
Our analysis is too simplified to draw the phase boundary (it is not important for the present paper). 
We are however able to discriminate the three different phases in the selected points 
indicated by the symbols in Fig.~\ref{appendix.fig:phase.diagram}. 
These values of the couplings have been used for the analysis of the two-site entanglement.

\section{Weak interaction limit} 
\label{weak}
In the regime in which the interaction $V$ in Eq.~\eqref{hamiltonian} is finite but small, 
typically of the order of $10^{-2} - 10^{-1}$, we can clearly separate two different time scales. 
At the first one ($t^\star$, introduced in Sec.~\ref{results}), 
the two-site entanglement saturates into a plateau, like in the AL phase.
After the second one (defined by $t_{\rm int}$), the power-law decay typical of the MBL phase begins. 
In this way we can set a clear distinction between AL and MBL regimes in the same time window. 
We expect the effect of MBL dynamics to appear at times of the order of $t_{\rm int} \sim 1/V$; 
for small interactions this scale can be much larger than the typical time scale associated 
to saturation in the AL phase: $t_{\rm int}\gg t^\star$.

\begin{figure}[t!]
  \includegraphics[scale=0.4]{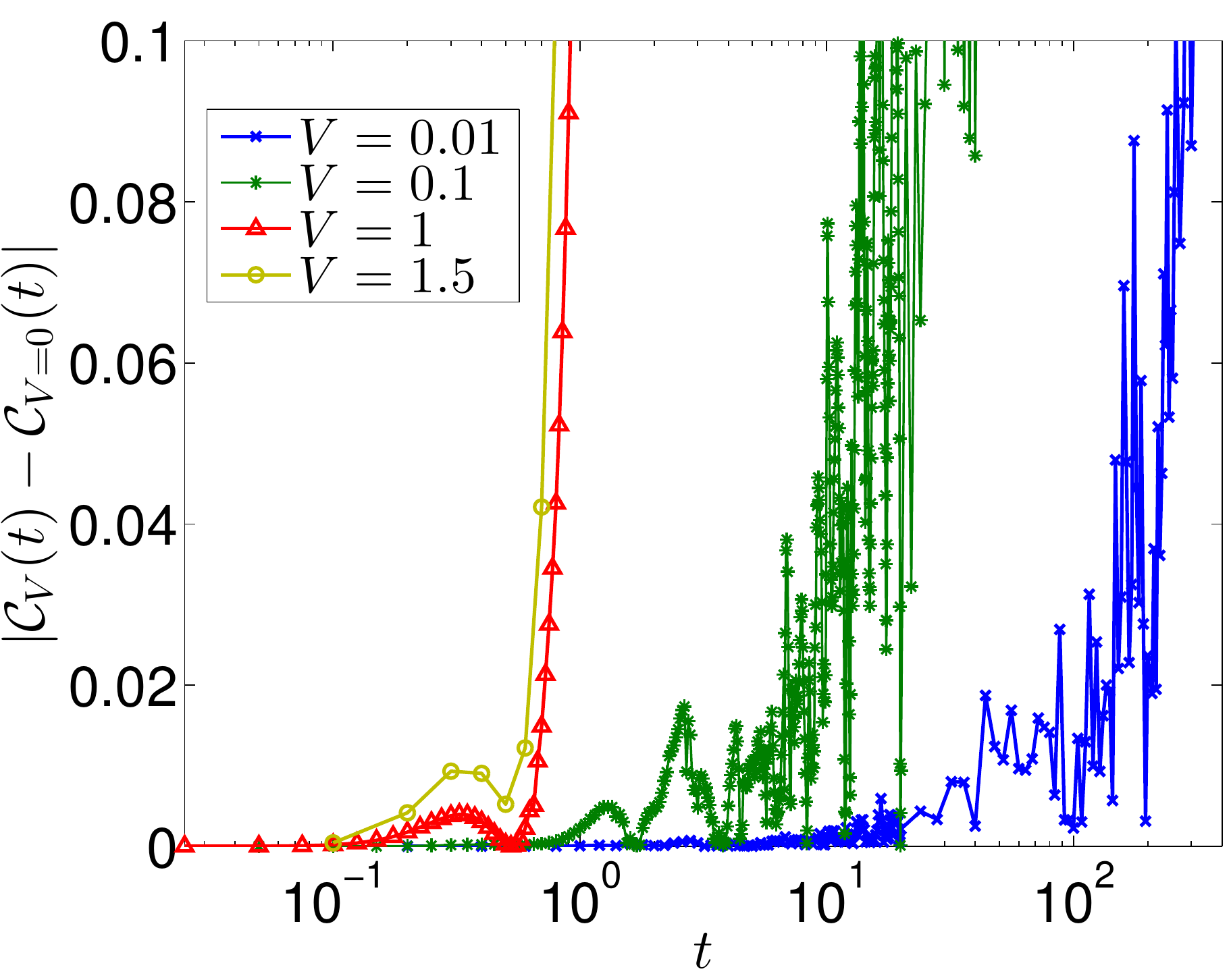}
  \includegraphics[scale=0.4]{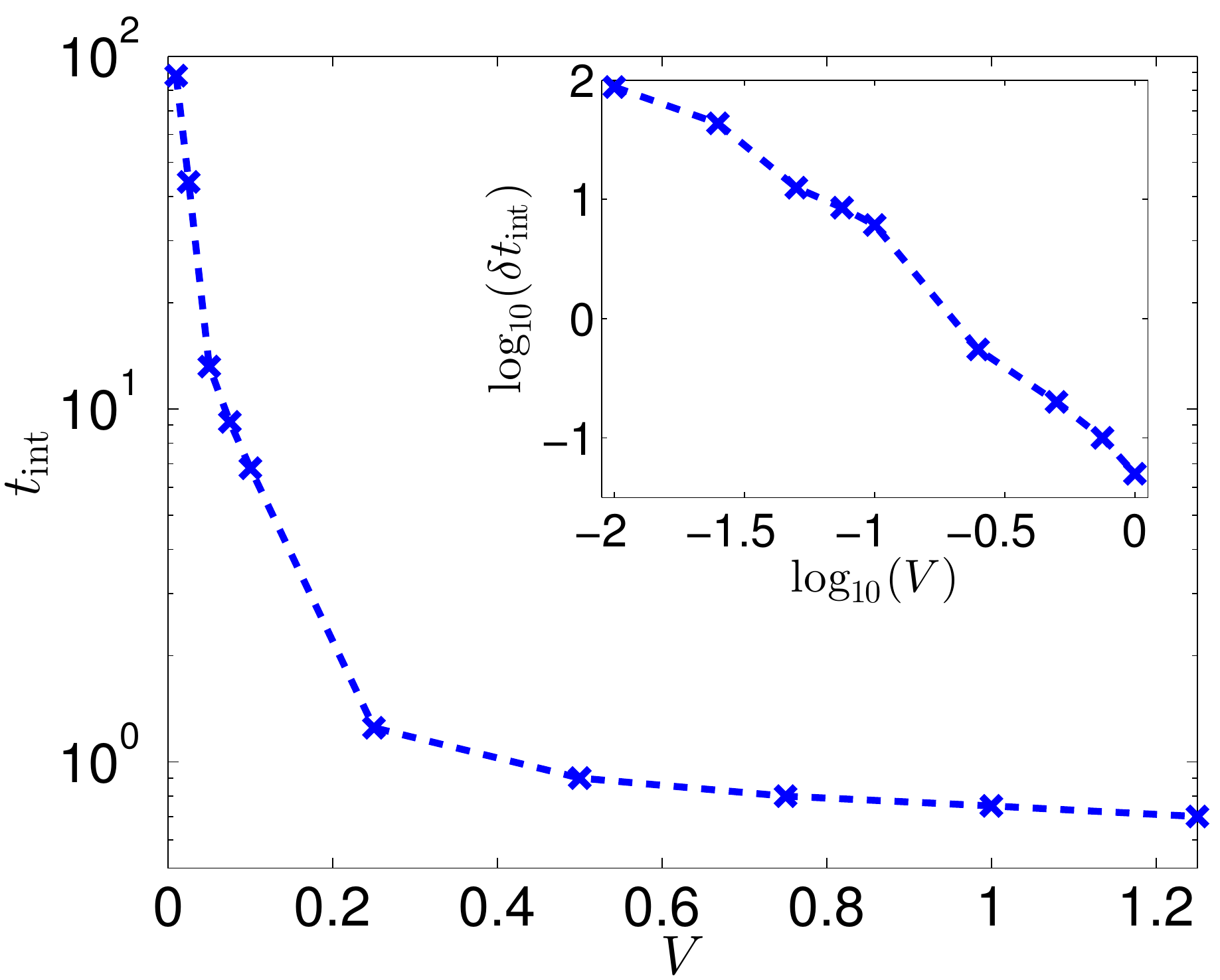}
  \caption{(Top) Difference $|{\cal C}_{V \ne 0}(t) - {\cal C}_{V=0}(t)|$ 
    as a function of time. The system has $L=24$ sites, fixed disorder strength $\Delta=3$, 
    and results are averaged over $\sim 30$ disorder realisations. The difference remains negligible 
    up to a characteristic time $t_{\rm int}$, when interactions become relevant. 
    Such time can be extrapolated by analysing when the curves start growing.
    (Bottom) Dependence of $t_{\rm int}$ with $V$. The inset displays
    $\delta t_{\rm int} \equiv t_{\rm int}(V,\Delta) - t_{\rm int}(V\gg 1,\Delta)$, 
    where $t_{\rm int}(V\gg 1,\Delta) \simeq 0.7$.}
  \label{concurrence-plat:fig}
\end{figure}

Our expectations are confirmed by the results shown in Fig.~\ref{concurrence-plat:fig}.
As we can see in the top panel, for very small values of $V$ the concurrence presents 
a ``plateau'' after the initial dynamics, where it is indistinguishable 
from the non-interacting case. Only after some finite time $t_{\rm int}$, the effects 
of interactions play a relevant role in the dynamics and the concurrence starts to decay. 
Since it is strictly related to the effect of interactions, we dub $t_{\rm int}$ as the interaction time.

We can give an estimate of $t_{\rm int}$ by extracting it from the time evolution of the concurrence. 
In order to do that, we start giving a more quantitative definition of $t_{\rm int}$. 
To this end it is illuminating to plot, at fixed pseudo-disorder strength $\Delta$, 
the difference $|{\cal C}_V(t) - {\cal C}_{V=0}(t)|$ (Fig.~\ref{concurrence-plat:fig}, top panel). 
The interaction time $t_{\rm int}$ is defined as the time at which the concurrence with $V>0$ 
starts to differ from the non-interacting one ($V=0$) more than a given threshold $\varepsilon$. 
We choose $\varepsilon = 0.025$, in such a way to capture the effect of the interactions, 
and not just small oscillations around the non interacting dynamics. 
The results however are qualitatively similar for slightly different values of $\varepsilon$. 

The result of this analysis is reported in the bottom panel of Fig.~\ref{concurrence-plat:fig}.
For the case shown in the figure, with disorder strength $\Delta = 3$, such interaction time 
corresponds to $t_{\rm int}(V) \propto V^{-a} + b$, with $a \approx 1.6$ and $b=0.7$ (see the inset). 
Therefore, our expectation of a $t_{\rm int}$ diverging as a power law for $V\to 0$
is confirmed, but the power-law exponent is different from what we expected.

\section{Correlations in the $\ell$-bit model} 
\label{corr-lbit}
This appendix summarises the derivation of the time dependent spin-spin correlation functions 
in the effective $\ell$-bit model~\eqref{ell.bit.model}: these correlations are necessary 
to determine the two-spin reduced density matrix and then the two-site entanglement 
(see also Ref.~[\onlinecite{Abanin.2014}]).
According to the $\ell$-bit phenomenological model, the physical spin operators 
$\{\hat{S}_j^\alpha\}_{j=1,\,\ldots,\,L}^{\alpha=\mathbb{I},x,y,z}$ of Hamiltonian~\eqref{hamiltonian} are,
in principle, ``locally'' related to the $\ell$-bit operators 
$\{\hat{\tau}_{m}^\alpha\}_{m=1,\,\ldots,\,L}^{\alpha=\mathbb{I},x,y,z}$ as
\begin{equation}
  \hat S_j^{\alpha} = \sum_m \sum_{\alpha'} \Delta_{j,m}^{\alpha,\alpha'} \hat \tau_m^{\alpha'} \,,
\end{equation}
where $\Delta_{j,m}^{\alpha,\alpha'}$ is localised, in the sense that it is non-vanishing only in a finite range 
around $m$. The precise form of $\Delta_{j,m}^{\alpha,\alpha'}$ is generally non-trivial to obtain, 
but this is not important for us. The main message is that we expect the properties 
of the $\ell$-bit operators and those of the physical spins to be similar.
This is confirmed by the results shown in Sec.~\ref{lbit-model}: 
the concurrence obtained from the correlations of the $\ell$-bit operators shows the same 
polynomial decay as the one numerically computed for the physical spins in Sec.~\ref{results}.
In this appendix we discuss in detail the computation of the dynamics of concurrence 
for pairs of $\ell$-bit sites, that has been considered in Sec.~\ref{lbit-model}.

For such task we first need to compute the reduced density matrix ${\rho}_{m,n}(t)$
for two different sites $n\neq m$,
which can be obtained from the correlations of its $\ell$-bit operators
\begin{equation}
  \rho_{m,n}(t) = \sum_{\alpha,\alpha'=\mathbb{I},x,y,z} \langle \hat{\tau}_m^{\alpha}(t)\,\hat{\tau}_n^{\alpha'}(t) \rangle
  \, \hat{\tau}_m^{\alpha}(0)\,\hat{\tau}_n^{\alpha'}(0)\,.
\end{equation}
It is important to stress that all the anlytical formulae we will find are valid for $n\neq m$.
Let us consider general initial uncorrelated states
\begin{eqnarray}
  && \ket{\psi(0)} = \otimes_{m=1}^L \ket{\chi_m(\phi_m,\theta_m)} \nonumber \\
  && \ket{\chi_m(\phi_m,\theta_m)} = \cos(\phi_m) \ket{\uparrow} + e^{i\theta_m}\sin(\phi_m)\ket{\downarrow}\,, \nonumber
\end{eqnarray}
where $\{\ket{\uparrow},\ket{\downarrow}\}$ are the eigenstates of the $\hat \tau_m^z$ operator. 
In this way, the correlators of any $\ell$-bit operators can be easily handled: evaluating
the expectations over the initial state, we find
\begin{equation}
  \langle \hat\tau_m^{\alpha}\hat\tau_n^{\alpha'}\rangle = \langle \hat\tau_m^{\alpha} \rangle \langle \hat\tau_n^{\alpha'} \rangle \,,
\end{equation}
due to the separability of this state. Exploiting this relation and using the Heisenberg representation,
all the correlators at any times can be analytically computed, as we will show below.

Since all the operators in the $\ell$-bit model commute, it is possible 
to analytically compute the time evolution of any operator in the Heisenberg picture. 
Since we will extensively use them in our analysis,
let us just briefly recall the commutation relation between the Pauli matrices:
\begin{equation}
  \big[ \hat \tau_j^\alpha, \hat \tau_j^\beta \big] = 2 i \, \epsilon_{\alpha \beta \gamma} \, \hat \tau_j^\gamma \,, \nonumber
\end{equation}
$\epsilon_{\alpha \beta \gamma}$ being the Levi-Civita coefficient. 
The $\ell$-bit operators evolve in the Heisenberg picture as
\begin{equation}
  \frac{d}{dt} \hat \tau_m^\alpha = i \big[ \hat H,\hat \tau_m^{\alpha} \big] \,. \nonumber
\end{equation}
Since $\hat \tau_m^{z}$ commute with the Hamiltonian $\hat{H}$ of Eq.~\eqref{ell.bit.model}, 
they are time independent. Let us focus on the $\alpha \neq z$ cases. 
Applying the commutation relations in the Heisenberg equations, we obtain
\begin{equation}
  \frac{d}{dt} \hat{\tau}_m^\alpha = \epsilon_{z\alpha \bar{\alpha} } \, \hat \tau_m^{\bar{\alpha}} \, \hat{A}_m \,,
  \label{first.time.der}
\end{equation}
with $\bar{\alpha} = y (x)$ for $\alpha = x (y)$, and 
\begin{equation}
  \hat{A}_m = -2 \bigg( h_m \mathbb{I} + 2 \sum_{j \neq m} \mathcal{J}_{m j} \hat \tau_j^{z} \bigg)\,. \nonumber
\end{equation}
The solution can be cast in the form
\begin{equation}
  \hat \tau_m^{\alpha}(t) = \hat{C}_m^{\alpha,-} e^{i\hat A_m t} + \hat{C}_m^{\alpha,+} e^{-i\hat A_m t},
\end{equation}
where 
\begin{equation}
  \hat C_m^{\alpha,\pm} = \tfrac12 \Big[ \hat \tau_m^{\alpha}(0) \pm i \, \epsilon_{z\alpha \bar{\alpha}} \, \hat \tau_m^{\bar{\alpha}}(0) \Big] \,. \nonumber
\end{equation}
Given the above expressions, we can explicitly compute the expectation values for all local 
and two-point correlations. We will extensively use the following identity
\begin{equation}
  e^{i\hat A_m t} = e^{-2 i h_m \mathbb{I} \, t} \,e^{-4 i \mathcal{J}_{mn} \hat \tau_n^z t} 
  \bigg( \! \prod_{j\neq m,n} e^{-4 i \mathcal{J}_{mj} \hat \tau_j^z t} \! \bigg) ,
\end{equation}
with $n$ being an arbitrary site index.
 
\noindent 
\textbf{Local averages.} 
For the $z$-spin terms we have 
\begin{equation}
  \langle \tau_m^{z}(t) \rangle = \langle \tau_m^{z}(0) \rangle = \cos^2(\phi_m) - \sin^2(\phi_m) \,.
\end{equation}
For the $x$-spin terms we obtain
\begin{align}
  & \langle \tau_m^{x}(t) \rangle = \sin(\phi_m) \cos(\phi_m) \left[ e^{-i\theta_m } e^{- 2 i h_m t} 
    K_{m,m}(t) + {\rm H.c.} \nonumber \right] ,\\ 
  & \mbox{with }
  K_{m,n}(t) = \prod_{j\neq m} \left( e^{-4i\mathcal{J}_{nj} t} \cos^2(\phi_j) + e^{4i\mathcal{J}_{nj} t} \sin^2(\phi_j) \right)\,. \nonumber
\end{align}
An analogous expression holds for the $y$-component.

\noindent
\textbf{Two-point correlations.} 
For the $zz$-correlations we have 
\begin{equation}
  \langle \hat \tau_m^z(t) \hat \tau_n^{z}(t) \rangle = \langle \hat \tau_m^z(0) \hat \tau_n^{z}(0) \rangle 
  = \langle \hat \tau_m^z(0) \rangle \langle \hat \tau_n^z(0) \rangle \,. \nonumber
\end{equation}
As for the $zx$-spin terms we obtain:
\begin{widetext}
\begin{equation}
  \langle \hat \tau_m^z(t) \hat \tau_n^{x}(t) \rangle = \Big[ e^{-4i\mathcal{J}_{nm} t} \cos^2(\phi_m) - e^{4i\mathcal{J}_{nm} t} \sin^2(\phi_m) \Big]
  e^{-i\theta_n}\sin(\phi_n) \cos(\phi_n) e^{-2 i h_n t} K_{m,n}(t) \, + \, {\rm H.c.} \,,
\end{equation}
where we used the fact that
\begin{equation}
  \langle \hat \tau_m^z e^{-4i \mathcal{J}_{nm} \hat \tau_m^z t} \rangle = 
  e^{-4i \mathcal{J}_{nm} t} \cos^2(\phi_m) - e^{4i\mathcal{J}_{nm} t} \sin^2(\phi_m) \,. \nonumber
\end{equation}
For the $zy$-spin terms we have
\begin{equation}
  \langle \hat \tau_m^z(t) \hat \tau_n^{y}(t) \rangle = i \Big[e^{-4i\mathcal{J}_{nm} t} \cos^2(\phi_m) - e^{4i\mathcal{J}_{nm} t} \sin^2(\phi_m) \Big] 
  e^{-i\theta_n}\sin(\phi_n) \cos(\phi_n) e^{-2ih_n t} K_{m,n}(t) \, + \, {\rm H.c.} \,. 
\end{equation}
Finally we compute the last correlation terms 
$\langle \hat \tau_m^\alpha(t) \, \hat \tau_n^{\alpha'}(t) \rangle $ (with $\alpha, \alpha' = x,y$),
which can be written as
\begin{equation}
  \hat \tau_m^{\alpha}(t) \hat \tau_n^{\alpha'}(t) = \sum_{b_m,b_n=-1,1} \hat X_{m,n}^{(\alpha,b_m)(\alpha',b_n)} \,,
  \label{appendix.two.point.corr}
\end{equation}
\begin{eqnarray}
  \mbox{with } \; \hat X_{m,n}^{(\alpha,b_m)(\alpha',b_n)} &=& \hat C_m^{\alpha,-b_m} \, e^{b_m i\hat A_m t} \, \hat C_n^{\alpha',-b_n} \, e^{b_n i\hat A_n t} \nonumber \\
  &=&
  \underbrace{\hat C_m^{\alpha,-b_m} \, e^{-2i(b_m h_m \mathbb{I} + b_n 2\mathcal{J}_{nm} \hat \tau_m^z) t}}_{\mbox{$m$ site}} \,
  \underbrace{e^{-2i(b_n h_n \mathbb{I} + b_m 2\mathcal{J}_{mn} \hat \tau_n^z) t} \, \hat C_n^{\alpha',-b_n} }_{\mbox{$n$ site}}\,
  \underbrace{\bigg( \prod_{j\neq m,n} e^{-4i(b_m \mathcal{J}_{mj}+ b_n \mathcal{J}_{nj})\hat \tau_j^z t}\bigg)}_{\mbox{rest}}\,. 
  \nonumber
\end{eqnarray}
The expectation value $\langle \hat X_{m,n}^{(\alpha,b_m)(\alpha',b_n)} \rangle$ is thus given by
\begin{equation}
  \langle \hat X_{m,n}^{(\alpha,b_m)(\alpha',b_n)} \rangle
  = e^{-2i(b_m h_m + b_n h_n ) t} \,G_t^\alpha(m,n,b_m,b_n) \, G_t^{\alpha'}(n,m,-b_n,-b_m)^* \, F_t(m,n,b_m,b_n) \,,
  \label{appendix.eq.Xcorr}
\end{equation}
\begin{eqnarray}
  \mbox{where } \qquad 
  F_t(m,n,b_m,b_n) \equiv \prod_{j\neq m,n} \big[ e^{-4i(b_m \mathcal{J}_{mj}+ b_n \mathcal{J}_{nj}) t} \cos^2(\phi_j) 
+ e^{4i(b_m \mathcal{J}_{mj} + b_n \mathcal{J}_{nj}) t} \sin^2(\phi_j) \big] \nonumber
\end{eqnarray}
and
\begin{eqnarray}
  G_t^{x}(m,n,-1,b_n) & = & e^{4i b_n \mathcal{J}_{nm} t} \, e^{ i\theta_m} \sin(\phi_m)\cos(\phi_m) \,, \nonumber \\
  G_t^{x}(m,n,+1,b_n) & = & e^{-4i b_n \mathcal{J}_{nm} t} \, e^{- i\theta_m} \sin(\phi_m)\cos(\phi_m) \,, \nonumber \\
  G_t^{y}(m,n,-1,b_n) & = & -i \langle \hat C_m^{x,+} e^{-4i b_n \mathcal{J}_{nm} \hat \tau_m^z t} \rangle \,, \nonumber\\
  G_t^{y}(m,n,+1,b_n) & = & i \langle \hat C_m^{x,-} e^{-4i b_n \mathcal{J}_{nm} \hat \tau_m^z t} \rangle \,. \nonumber
\end{eqnarray}
\end{widetext}
Using Eqs.~\eqref{appendix.two.point.corr} and~\eqref{appendix.eq.Xcorr}, 
it is possible to compute $\langle \hat \tau_m^{\alpha=x,y}(t) \hat \tau_n^{\alpha'=x,y}(t) \rangle$.

\section{Finite-size corrections and errors due to statistical averages}
\label{finitesize}

\begin{figure}[!b]
  \includegraphics[width=7.7cm]{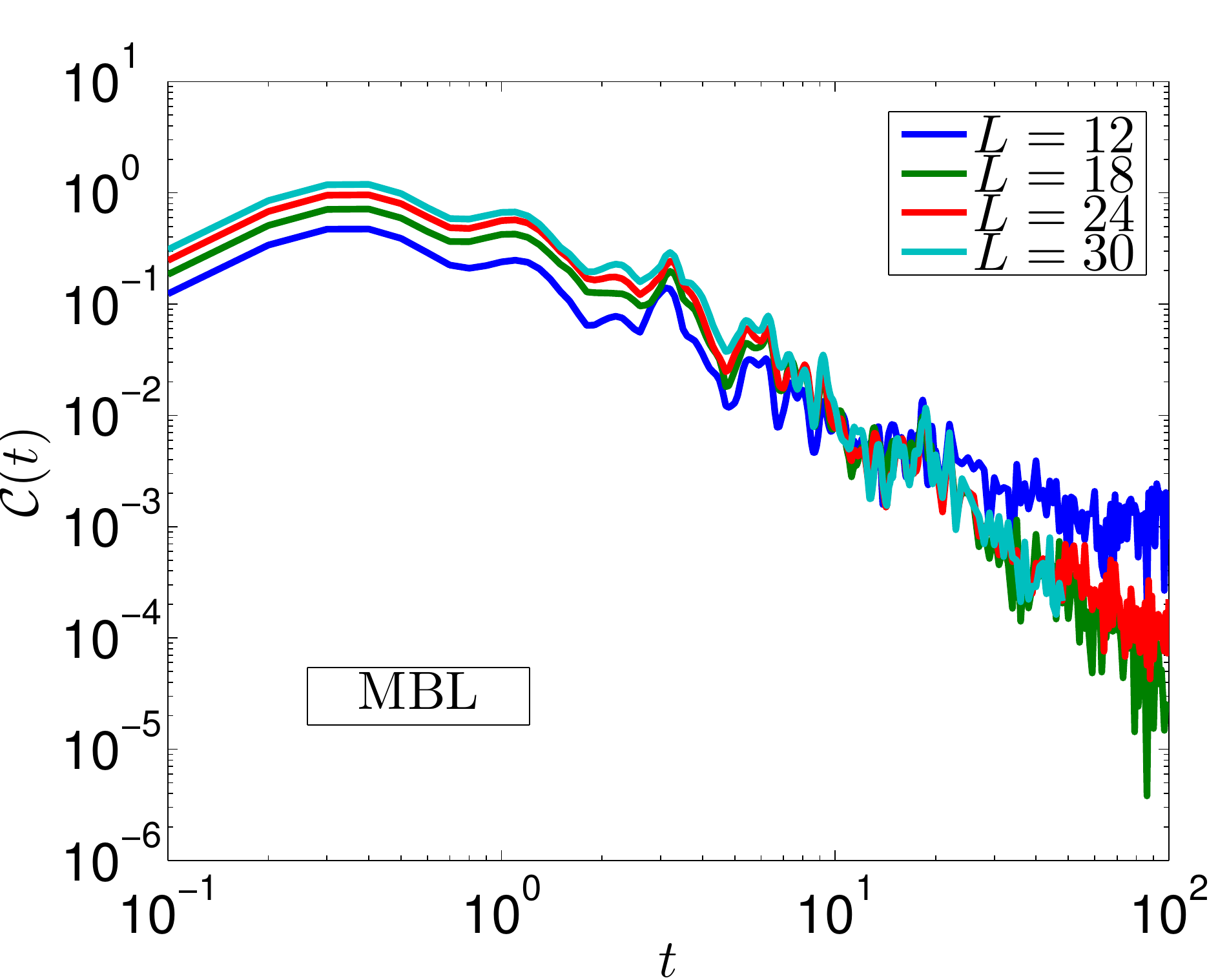}
  \includegraphics[width=7.5cm]{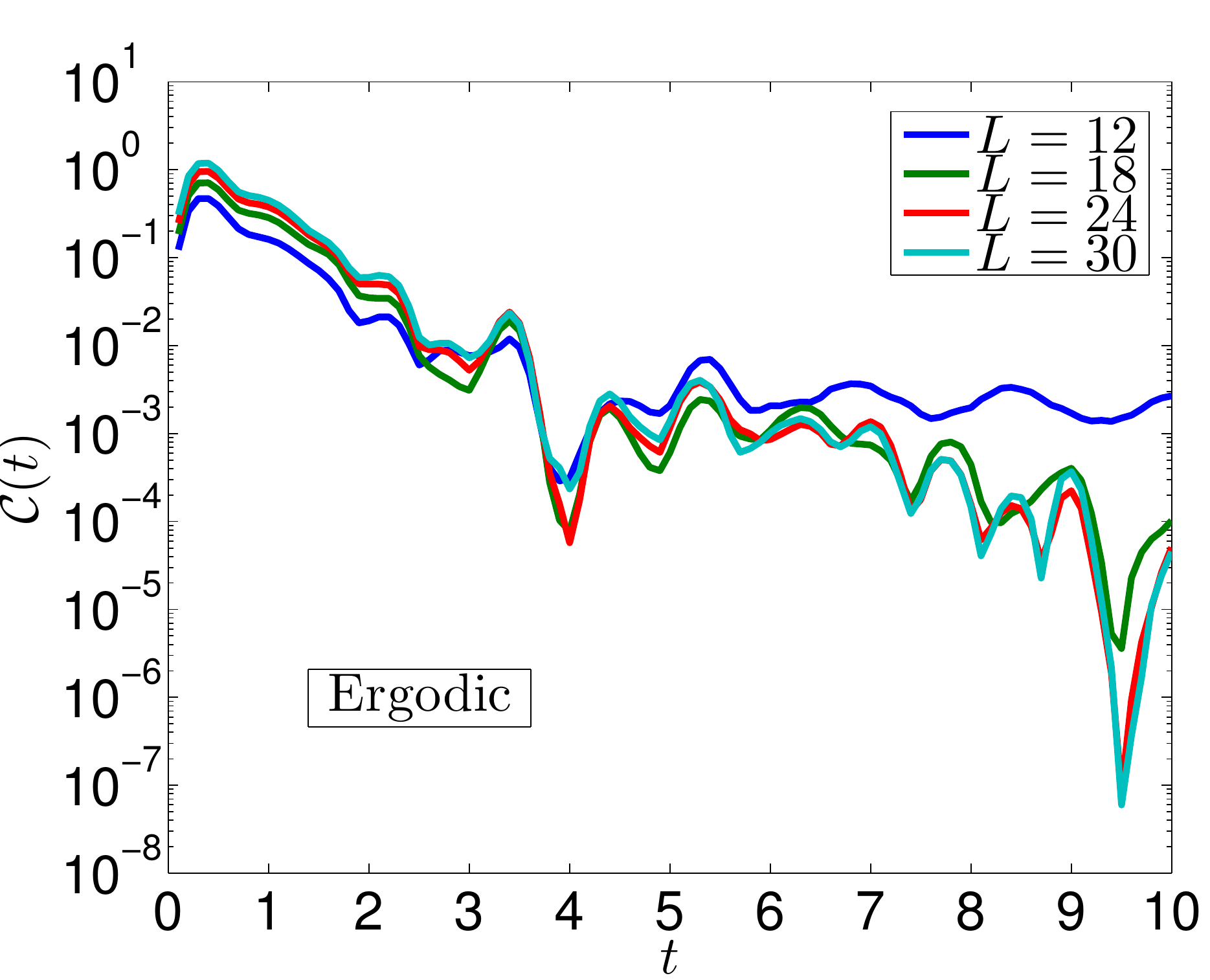}
  \caption{Time evolution for the concurrence in the MBL phase (top panel,
    $\Delta=3$, $V=1$) and in the ergodic phase (bottom panel, $\Delta=3$, $V=1.5$).
    The various data sets are for different system sizes, according to the legend.}
  \label{conc-fss}
\end{figure}

\begin{figure}[!b]
  \includegraphics[width=7.5cm]{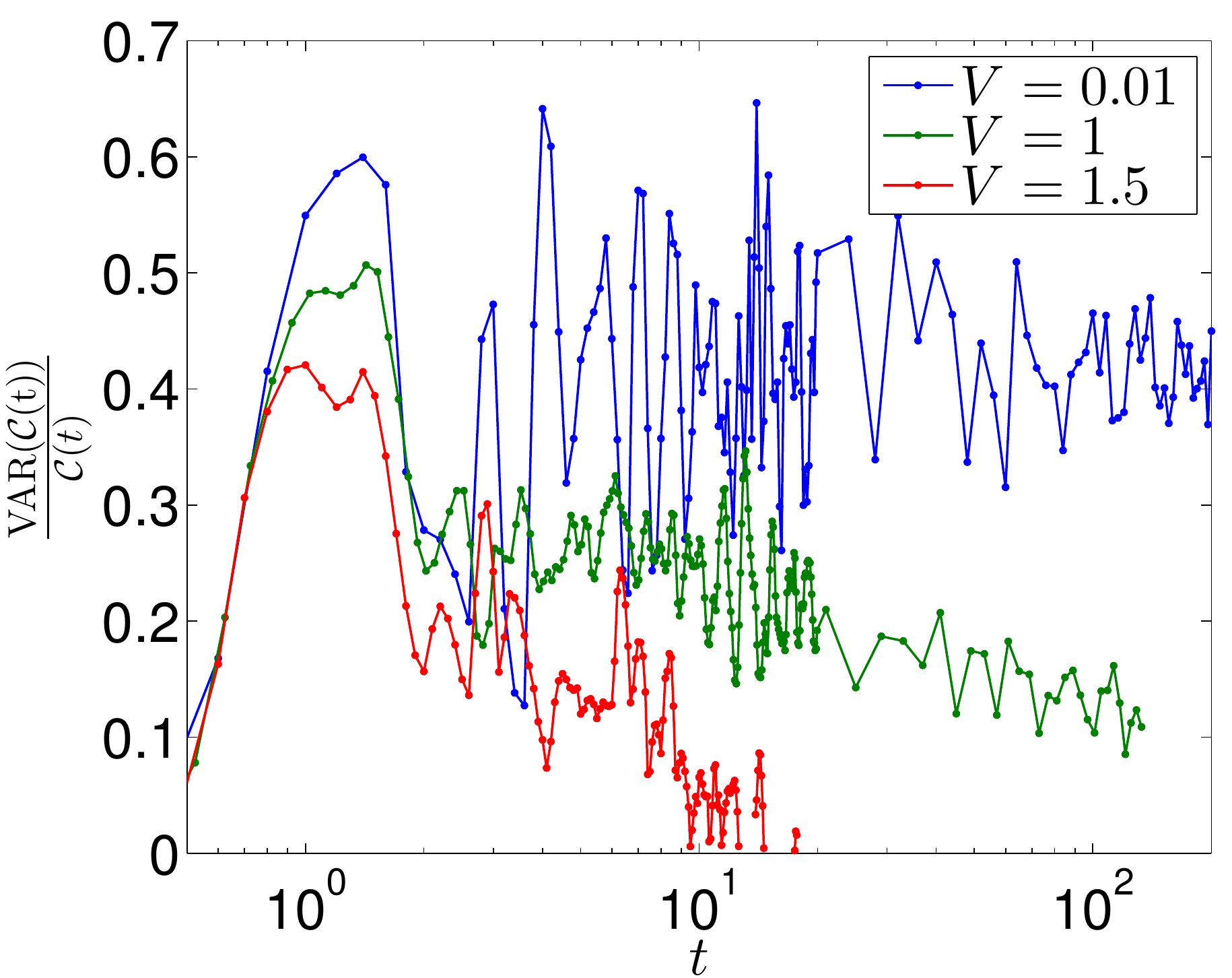}
  \includegraphics[width=7.5cm]{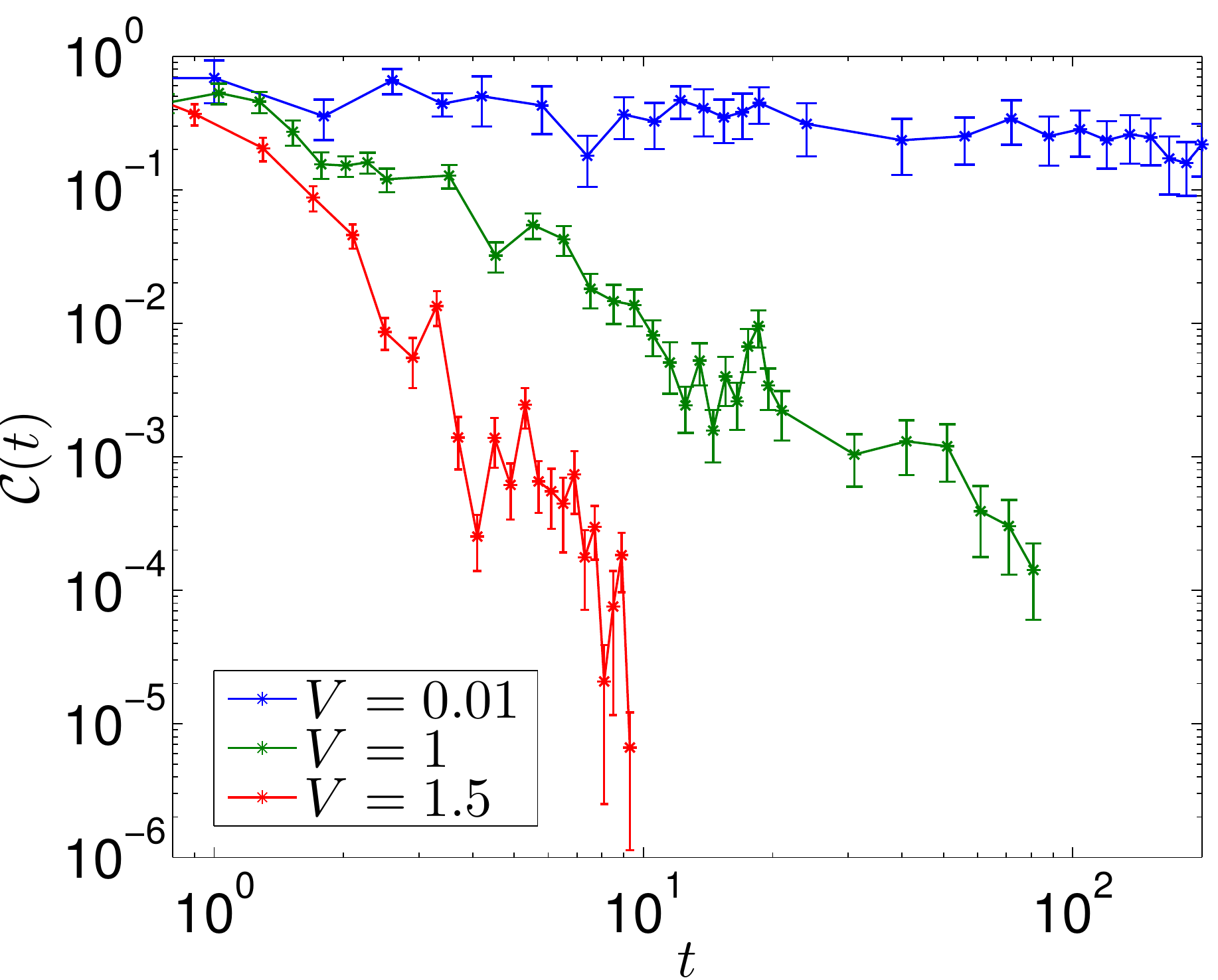}
  \caption{(Top) Relative variance of the concurrence as a function of time,
    as quantified by Eq.~\eqref{eq:concvar}.
    The data refer to three emblematic cases for $\Delta=3$, in the ergodic phase
    ($V=1,5$), the MBL phase ($V=1$), and very close to the AL phase ($V=0.01$).
    (Bottom) Concurrence as a function of time, for the same cases as before.
    Error bars quantify the variance with respect to the disorder average.
    The results shown in the plots are averaged over $30$ pseudo-disorder realizations.}
  \label{conc-variance}
\end{figure}

We first discuss the influence that finite-size effects may have during
the relaxation dynamics of model~\eqref{fermion.model}.
Due to the presence of disorder, which is the responsible of many-body dephasing,
we expect that deep in the localized phase the relaxation process
will not be very sensitive to the size of the simulated systems, 
up to the time scales we are able to reach. 
This is what we observe in Fig.~\ref{conc-fss}, when monitoring 
the time evolution of the concurrence ${\cal C}$.
Finite-size effects in the MBL phase start to be visible for $L=12$
sites at $t \gtrsim 10$, while they are are virtually absent
on the same time scale for $L\geq 18$ (top panel).
In the ergodic phase, due to the absence of localization,
one would expect a more pronounced size dependence.
Anyway in view of the very small time scales we are able to reach ($t \lesssim 10$),
in all our simulations finite-size corrections
are safely under control for $L \geq 18$ (bottom panel).
We point out that, in order to avoid any unwanted dependence 
on the matrix truncation of our MPS simulations, 
we fixed the bond link at $m=200$ for all the data sets.
For such value of $m$, we carefully checked that our results have reached the convergence
tor the times that are plotted in the figures.

We conclude this section by discussing the variance of our quantities of interest,
which is induced by the averages over different disorder realizations.
Let us focus on the average concurrence defined in Eq.~\eqref{conc:eqn} and calculate
its variance by propagating the error according to
\begin{eqnarray}
  {\rm Var} \big( {\cal C}(t) \big) &=& \sum_{i,j\in \rm{bulk}} \bigg\vert \frac{\partial {\cal C}(t)}{\partial C_{ij}} \bigg\vert  {\rm Var} \big( C_{ij}(t) \big) \nonumber \\
  &=& \sum_{i,j\in \rm{bulk}}  2 \, {\cal C}(t) \, {\rm Var} \big( C_{ij}(t) \big) \,, 
  \label{eq:concvar}
\end{eqnarray}
where ${\rm Var} \big( C_{ij}(t) \big)$ is the variance of the two-site concurrence. 
The data in Fig.~\ref{conc-variance} display three representative cases,
where it emerges that the relative error induced
by the pseudo-disorder avereages (top panel) is decreasing with $V$.
The averages have been performed over the same number of pseudo-disorder realizations
as the data presented in Sec.~\ref{results}.
Apart from the quasi-AL phase (where we were however able to substantially increase the statistics,
due to the integrability of the model), we notice that it also decreases with time,
in such a way that our description for the concurrence long-time dynamics
is basically unaffected (bottom panel).
A similar result can be found for the imbalance ${\cal I}$ of Eq.~\eqref{imbalance}
(data not shown here).


\begin{thebibliography}{100}

\bibitem{basko2006metal}
  D.~Basko, I.~Aleiner, and B.~Altshuler,
  \newblock Ann. Phys. {\bf 321}, 1126 (2006).

\bibitem{oganesyan2007localization}
  V.~Oganesyan and D.~A.~Huse,
  \newblock Phys. Rev. B {\bf 75}, 155111 (2007).

\bibitem{Nandkishore:2015aa}
  R.~Nandkishore and D.~A.~Huse,
  \newblock Annu. Rev. Condens. Matter Phys. {\bf 6}, 15 (2015).

\bibitem{Berry_LH84}
  M.~V.~Berry,
  \newblock {Semiclassical mechanics of regular and irregular motion},
  \newblock in {\em {Les Houches, Session XXXVI, 1981 --- Chaotic Behaviour of Deterministic Systems}}, 
  edited by R.~S.~G.~Ioos, R.~H.~G.~Helleman, pp.~174-271 (North-Holland Publishing Company, 1983).

\bibitem{Arnold:book}
  V.~I.~Arnol{'}d,
\newblock {\em {Mathematical Methods of Classical Mechanics}} (Springer, 1989).

\bibitem{Polkovnikov_RMP11}
  A.~Polkovnikov, K.~Sengupta, A.~Silva, and M.~Vengalattore,
  \newblock Rev. Mod. Phys. {\bf 83}, 863 (2011).

\bibitem{Abanin.2103.LIOM} 
  M.~Serbyn, Z.~Papic, and D.~A.~Abanin,
  \newblock Phys. Rev. Lett. {\bf 111}, 127201 (2013).

\bibitem{ros2015integrals}
  V.~Ros, M.~Mueller, and A.~Scardicchio,
  \newblock Nucl. Phys. B {\bf 891}, 420 (2015).

\bibitem{Abanin_PRB}
  A.~Chandran, I.~H.~Kim, G.~Vidal, and D.~A.~Abanin,
  \newblock Phys. Rev. B {\bf 91}, 085425 (2015).

\bibitem{imbrie2014many}
  J.~Z.~Imbrie,
  \newblock J. Stat. Phys. {\bf 163}, 998 (2016).

\bibitem{Pasqualazzo}
  P.~Calabrese, F.~H.~L.~Essler, and G.~Mussardo,
  \newblock J. Stat. Mech. {\bf 064001}, (2016).

\bibitem{Mauro_PRB}
  M.~Schiulaz, A.~Silva, and M.~M{\"u}ller,
  \newblock Phys. Rev. B {\bf 91}, 184202 (2015).

\bibitem{Pino_arXiv15}
  M.~Pino, B.~L.~Altshuler, and L.~B.~Ioffe,
  \newblock Proc. Natl. Acad. Sci. USA {\bf 113}, 536 (2016).

\bibitem{anderson1958absence}
  P.~W.~Anderson,
  \newblock Phys. Rev. {\bf 109}, 1492 (1958).

\bibitem{Abrahams2010}
  E.~Abrahams {(editor)},
  \newblock {\em 50 years of Anderson localization} (World Scientific, 2010).

\bibitem{pal2010mb}
  A.~Pal and D.~A.~Huse,
  \newblock Phys. Rev. B {\bf 82}, 174411 (2010).

\bibitem{Luitz:2016aa}
  D.~J.~Luitz, N.~Laflorencie, and F.~Alet,
  \newblock Phys. Rev. B {\bf 93}, 060201 (2016).

\bibitem{PhysRevLett.117.040601}
  M.~\v{Z}nidari\v{c}, A.~Scardicchio, and V.~K.~Varma,
  \newblock Phys. Rev. Lett. {\bf 117}, 040601 (2016).

\bibitem{Kerala-Varma:2015aa}
  V.~K.~Varma, A.~Lerose, F.~Pietracaprina, J.~Goold, and A.~Scardicchio,
  \newblock arXiv:1511.09144 (2015).

\bibitem{huse2014phenomenology}
  D.~A.~Huse, R.~Nandkishore, and V.~Oganesyan,
  \newblock Phys. Rev. B {\bf 90}, 174202 (2014).

\bibitem{znidaric2008many}
  M.~\v{Z}nidari\v{c}, T.~Prosen, and P.~Prelov\v{s}ek,
  \newblock Phys. Rev. B {\bf 77}, 064426 (2008).

\bibitem{Bardason2012}
  J.~H.~Bardarson, F.~Pollmann, and J.~E.~Moore,
  \newblock Phys. Rev. Lett. {\bf 109}, 017202 (2012).

\bibitem{Abanin.2013.EntGrowth} 
  M.~Serbyn, Z.~Papic, and D.~A.~Abanin,
  \newblock Phys. Rev. Lett. {\bf 110}, 260601 (2013).

\bibitem{de2013ergodicity}
  A.~De~Luca and A.~Scardicchio,
  \newblock Europhys. Lett. {\bf 101}, 37003 (2013).

\bibitem{Monthus:2016aa}
  C.~Monthus,
  \newblock J. Stat. Mech. (2016) P04005 (2016).

\bibitem{Canovi_PRB11}
  E.~Canovi, D.~Rossini, R.~Fazio, G.~E.~Santoro, and A.~Silva,
  \newblock Phys. Rev. B {\bf 83}, 094431 (2011).

\bibitem{Abanin.2014} 
  M.~Serbyn, Z.~Papic, and D.~A.~Abanin,
  \newblock Phys. Rev. B {\bf 90}, 174302 (2014).

\bibitem{berkone}
  R.~Berkovits,
  \newblock Phys. Rev. Lett. {\bf 108}, 176803 (2012).

\bibitem{belone}
  B.~Bauer, and C.~Nayak,
  \newblock J. Stat. Mech. (2013) P09005.

\bibitem{Pal_Huse_PRB10}
  A.~Pal and D.~A.~Huse,
  \newblock Phys. Rev. B {\bf 82}, 174411 (2010).

\bibitem{huse2013localization}
  D.~A.~Huse, R.~Nandkishore, V.~Oganesyan, A.~Pal, and S.~L.~Sondhi,
  \newblock Phys. Rev. B {\bf 88}, 014206 (2013).

\bibitem{altman-review}
  E.~Altman and R.~Vosk,
  \newblock Annu. Rev. Condens. Matter Phys. {\bf 6}, 383 (2015).

\bibitem{Demler-interferometry}
  M.~Serbyn, M.~Knap, S.~Gopalakrishnan, Z.~Papi\'c, N.~Y.~Yao, C.~R.~Laumann, D.~A.~Abanin, M.~D.~Lukin, and E.~A.~Demler,
  \newblock Phys. Rev. Lett. {\bf 113}, 147204 (2014).

\bibitem{Moore-revivals}
  R.~Vasseur, S.~A.~Parameswaran, and J.~E.~Moore,
  \newblock Phys. Rev. B {\bf 91}, 140202 (2015).

\bibitem{schreiber2015observation}
  M.~Schreiber, S.~S.~Hodgman, P.~Bordia, H.~P.~L\"uschen, M.~H.~Fisher, R.~Vosk, E.~Altman, U.~Schneider, and I.~Bloch,
\newblock Science {\bf 349}, 842 (2015).

\bibitem{schn_MBL_per}
  P.~Bordia, H.~L{\"u}schen, U.~Schneider, M.~Knap, and I.~Bloch,
  \newblock arXiv:1607.07868 (2016).

\bibitem{Monroe_MBL}
  J.~Smith, A.~Lee, P.~Richerme, B.~Neyenhuis, P.~W.~Hess, P.~Hauke, M.~Heyl, D.~A.~Huse, and C.~Monroe,
  \newblock Nat. Phys. {\bf 12}, 907 (2016).

\bibitem{Fukuhara_ent}
  T.~Fukuhara, S.~Hild, J.~Zeiher, P.~Schau\ss, I.~Bloch, M.~Endres, and C.~Gross,
  \newblock Phys. Rev. Lett. {\bf 115}, 035302 (2015).

\bibitem{Jurcevic_ent}
  P.~Jurcevic, B.~P.~Lanyon, P.~Hauke, C.~Hempel, P.~Zoller, R.~Blatt, and C.~F.~Roos,
  \newblock Nature {\bf 511}, 202 (2014).

\bibitem{Dechiara_JSTAT06}
  G.~De~Chiara, S.~Montangero, P.~Calabrese, and R.~Fazio,
  \newblock J. Stat. Mech. (2006) P03001.

\bibitem{Deng_arX16}
  D.-L.~Deng, X.~Li, J.~H. Pixley, Y.-L.~Wu, and S.~D.~Sarma,
  \newblock arXiv:1607.08611 (2016).

\bibitem{Roosz2014}
 G.~Ro\'osz, U.~Divakaran, H.~Rieger, and F.~Igl\'oi,
  Phys. Rev. B {\bf 90}, 184202 (2014).

\bibitem{Igloi2012}
 F.~Igl\'oi, Z.~Szatm\'ari, and Y.-C.~Lin,
  Phys. Rev. B {\bf 85}, 094417 (2012).

\bibitem{Zhao2016}
  Y.~Zhao, F.~Andraschko, and J.~Sirker,
  Phys. Rev. B {\bf 93}, 205146 (2016).

\bibitem{Greiner_ent}
  A.~M. Kaufman, M.~E.~Tai, A.~Lukin, M.~Rispoli, R.~Schittko, P.~M.~Preiss, and M.~Greiner,
  \newblock Science {\bf 353}, 794 (2016).

\bibitem{Aubry-Andre}
  S.~Aubry and G.~Andr{\'e},
  \newblock Analyticity breaking and anderson localization in incommensurate lattices,
  \newblock in {\em Group theoretical methods in physics (Proc. Eighth Internat. Colloq., Kiryat Anavim, 1979)},
  Ann. Israel Phys. Soc. Vol.~3, pp.~133-164 (Hilger, Bristol, 1980).

\bibitem{kuklov_XXZ}
  A.~B.~Kuklov and B.~V.~Svistunov,
  \newblock Phys. Rev. Lett. {\bf 90}, 100401 (2003).

\bibitem{duan_XXZ}
  L.-M.~Duan, E.~Demler, and M.~D.~Lukin,
  \newblock Phys. Rev. Lett. {\bf 91}, 090402 (2003).

\bibitem{Mastropietro_PRL}
  V.~Mastropietro,
  \newblock Phys. Rev. Lett. {\bf 115}, 180401 (2015).

\bibitem{roati2008anderson}
  G.~Roati, C.~D'Errico, L.~Fallani, M.~Fattori, C.~Fort, M.~Zaccanti, G.~Modugno, M.~Modugno, and M.~Inguscio,
  \newblock Nature {\bf 453}, 895 (2008).

\bibitem{Woot_PRL}
  W.~K.~Wootters,
  \newblock Phys. Rev. Lett. {\bf 80}, 2245 (1998).

\bibitem{amico2008entanglement}
  L.~Amico, R.~Fazio, A.~Osterloh, and V.~Vedral,
  \newblock Rev. Mod. Phys. {\bf 80}, 517 (2008).

\bibitem{Coffman_tangle}
  V.~Coffman, J.~Kundu, and W.~K.~Wootters,
  \newblock Phys. Rev. A {\bf 61}, 052306 (2000).

\bibitem{Osborne_tangle}
  T.~J.~Osborne and F.~Verstraete,
  \newblock Phys. Rev. Lett. {\bf 96}, 220503 (2006).

\bibitem{Daley_JSTAT04}
  A.~J.~Daley, C.~Kollath, U.~{Schollw\"ock}, and G.~Vidal,
  \newblock J. Stat. Mech. (2004) P04005.

\bibitem{Schollwock_rev}
  U.~Schollwock,
  \newblock Ann. Phys. {\bf 326}, 96 (2011).

\bibitem{Bera_PRB}
  S.~Bera and A.~Lakshminarayan,
  \newblock Phys. Rev. B {\bf 93}, 134204 (2016).

\bibitem{Note1}
  The distribution of $\protect \mathcal {J}_{jl}$ for a local MBL model has long tails, 
  as found in Refs.~\onlinecite{ros2015integrals,pekker2015identifying}. 
  However, we found that this does not qualitatively change our results.

\bibitem{Mazza_NJP}
  L.~Mazza, D.~Rossini, R.~Fazio, and M.~Endres,
  \newblock New J. Phys. {\bf 17}, 013015 (2015).

\bibitem{Campbell.2016} 
  S.~Campbell, M.~J.~M.~Power, and G.~De~Chiara,
  \newblock arXiv:1608.08897 (2016).

\bibitem{buccheri2011structure}
  F.~Buccheri, A.~De~Luca, and A.~Scardicchio,
  \newblock Phys. Rev. B {\bf 84}, 094203 (2011).

\bibitem{luitz2016anomalous}
  D.~J.~Luitz and Y. BarLev,
  \newblock Phys. Rev. Lett. {\bf 117}, 170404 (2016).

\bibitem{Tomasi2016} 
  G.~De~Tomasi, S.~Bera, J.~H.~Bardarson, and F.~Pollmann,
  \newblock arXiv:1608.07183 (2016).

\bibitem{Haake:book}
  F.~Haake,
  \newblock {\em Quantum Signatures of Chaos ({$2^{\rm nd}$} ed.)} (Springer, 2001).

\bibitem{Berry_PRS76}
  M.~V.~Berry and M.~Tabor,
  \newblock Proc. Roy. Soc. {\bf A349}, 101 (1976).

\bibitem{Poilblanc_EPL93}
  D.~Poilblanc, T.~Ziman, J.~Bellisard, F.~Mila, and G.~Montambaux,
  \newblock Europhys. Lett. {\bf 22}, 537 (1993).

\bibitem{pekker2015identifying}
  D.~Pekker, B.~Tian, X.~Yu, B.~Clark, and V.~Oganesyan,
  \newblock Identifying the local conserved quantities in many-body-localized matter,
  \newblock in {\em APS Division of Atomic, Molecular and Optical Physics Meeting Abstracts} 
  Vol.~1, p.~6001P, 2015.

\end{thebibliography}
\end{document}